\documentclass[fleqn,usenatbib]{mnras}

\usepackage{newtxtext,newtxmath}
\usepackage[T1]{fontenc}

\usepackage{graphicx}	% Including figure files
\usepackage{amsmath}	% Advanced maths commands
\usepackage{mathrsfs}
\usepackage{hyperref}
\usepackage{pdflscape}
\usepackage{float}
\usepackage{here}
\usepackage[export]{adjustbox}
\usepackage{array}
\newcommand{\nustar}{\textit{NuSTAR }} 
\newcommand{\xmm}{\textit{XMM-Newton }} 
\newcommand{\logxi}{$\log(\xi / \rm erg\, \rm s^{-1}\,\rm cm)$}

\title[SEAMBHs. I. X-ray Continuum]{Systematic Broad-band X-ray Study of super-Eddington Accretion onto Supermassive Black Holes. I. X-ray Continuum}
\author[A. Tortosa et al.]{
Alessia Tortosa $^{1}$\thanks{E-mail: alessia.tortosa@mail.udp.cl},
Claudio Ricci $^{1,2,3}$,
Luis C. Ho $^{2,4}$,
Francesco Tombesi $^{5,6,7,8,9}$,
Pu Du $^{10}$,
Kohei Inayoshi $^{2}$,
\newauthor
Jian-Min Wang $^{10,11,12}$,
Jinyi Shangguan$^{13}$,
Ruancun Li$^{2}$
\\
$^{1}$Núcleo de Astronomía de la Facultad de Ingeniería, Universidad Diego Portales, Av. Ejército Libertador 441, Santiago, Chile\\
$^{2}$Kavli Institute for Astronomy and Astrophysics, Peking University, Beijing 100871, China\\
$^{3}$George Mason University, Department of Physics \& Astronomy, MS 3F3, 4400 University Drive, Fairfax, VA 22030, USA\\
$^{4}$ Department of Astronomy, School of Physics, Peking University, Beijing 100871, China\\
$^{5}$Dipartimento di Fisica, Univerisità di Roma Tor Vergata, via della Ricerca Scientifica 1, I-00133 Roma, Italy\\
$^{6}$INAF – Osservatorio Astronomico di Roma, Via Frascati 33, 00040 Monte Porzio Catone, Italy\\
$^{7}$ Department of Astronomy, University of Maryland, College Park, MD 20742, USA\\
$^{8}$ NASA Goddard Space Flight Center, Greenbelt, MD 20771, USA\\
$^{9}$ INFN - Roma Tor Vergata, Via della Ricerca Scientifica 1, 00133 Rome, Italy\\
$^{10}$ Key Laboratory for Particle Astrophysics, Institute of High Energy Physics, Chinese Academy of Sciences, 19B Yuquan Road, Beijing 100049, China\\
$^{11}$ School of Astronomy and Space Sciences, University of Chinese Academy of Sciences, 19A Yuquan road, Beijing 100049, China\\
$^{12}$ National Astronomical Observatory of China, 20A Datun Road, Beijing 100020, China\\
$^{13}$  Max-Planck Institute for Extraterrestrial Physics (MPE), Giessen-bachstr. 1, 85748 Garching, Germany
}

\date{Accepted XXX. Received YYY; in original form ZZZ}

\pubyear{2022}

\begin{document}
\label{firstpage}
\pagerange{\pageref{firstpage}--\pageref{lastpage}}
\maketitle

\begin{abstract}
We present the first systematic broad-band X-ray study of super-Eddington accretion onto SMBHs with simultaneous \nustar and \xmm or \textit{Swift}/XRT observations of a sample of 8 super-Eddington accreting AGN with Eddington ratio $1<\lambda_{\rm Edd}<426$. We find that the SEAMBHs show a steep primary continuum slope as expected for sources accreting in the super Eddington regime, mostly dominated by relativistic reflection. The Iron K$\alpha$ emission lines of the sources of our sample show relativistic broadening. In addition the equivalent widths of the narrow components of the Iron K$\alpha$ lines follow the 'X-ray Baldwin' effect, also known as the ‘Iwasawa-Taniguchi’ effect. We found a statistically significant correlation between the photon-index of the primary power-law and the Eddington ratio, consistent with past studies. Moreover, as expected for super-Eddington sources, the median value of the reflection fraction of the sources we analysed is a factor $\sim 2$ higher than the median reflection fraction value of the type\,1 AGN from the BASS sample. We are able to estimate the coronal temperature for three sources of our sample: Mrk\,382 ($kT_e=7.8$\,keV), PG\,0026+129 ($kT_e=39$\,keV) and IRAS\,04416+1215 ($kT_e=3$\,keV). Looking at the position of the SEAMBHs sources of our sample in the compactness-temperature diagram it appears that in super-Eddington AGN, as for lower Eddington ratio AGN, the X-ray corona is controlled by pair production and annihilation.
\end{abstract}


%%%%%%%%%%%%%%%%%%%%%%%%%%%%%%%%%%%%%%%%%%%%%%%%%%

%%%%%%%%%%%%%%%%% BODY OF PAPER %%%%%%%%%%%%%%%%%%

\section{Introduction}
Active Galactic Nuclei (AGN) are strong sources of X-ray radiation powered by the accretion of gas onto the central Supermassive Black Holes (SMBH, $M_{\rm BH } > 10^5 M_{\odot}$) \citep{Salpeter1964,2008ARA&A..46..475H}, which are found ubiquitously at the center of massive galaxies. The observed Eddington luminosity of most AGN is considered as an upper limit to the total luminosity that can be radiated by a compact object of mass $M_{\rm BH}$ in a situation of radiation-pressure equilibrium. Even if the observed bolometric luminosity of most AGN is smaller than the Eddington value ($L_{\rm Edd}=4\pi M_{\rm BH} m_p c/\sigma_T$, \citealt{1916MNRAS..77...16E}) \citep{Koratkar_1999,2011ApJ...728...98D,Netzer2013} a super-Eddington accretion rate, i.e. $L_{\rm bol}/L_{\rm Edd}=\lambda_{\rm Edd} > 1$ , is thought to be possible \citep{2012Sci...337..544V,2013ApJ...769...51Z}. The X-ray emission of AGN is produced in their innermost regions, surrounding the SMBH (few to tens of gravitational radii). It is due to the inverse-Compton scattering of the thermal UV/optical seeds photons, emitted from the accretion disk, by the coronal hot electrons. This process creates a X-ray power-law continuum (e.g., \citealt{1980A&A....86..121S}; \citealt{1993ApJ...413..507H}) which is the most direct probe for investigating AGN accretion properties.\\
In a situation of super-Eddington accretion, the nature of the accretion flow is expected to change dramatically by photon trapping through electron scattering in the dense matter and advection cooling \citep{1988ApJ...332..646A,Wang2014a}. The accretion disk can become 'slim', with the radiation being emitted anisotropically and with a large fraction of the overall energetic output being carried away in the form of outflows \citep{Jiang2019,Okuda_2021}. Supercritical accretion flows are expected to produce radiation-pressure driven outflows that will scatter soft X-rays photons from the underlying accretion flow via inverse Compton effect, producing the hard emission. From radiation-hydrodynamic simulations it can be shown that from a super-Eddington accretion flow the expected photon index of the primary power-law emission is expected to be $\sim 2$ \citep{King2003,2009PASJ...61..769K}.\\
The study of the hard X-ray emission of super-Eddington accreting black holes is extremely important to explain the fast growth of the first supermassive black holes in the early Universe (e.g., \citealt{begelman2006}), as well as to interpret the properties of tidal disruption events in which the mass accretion rate can exceed the Eddington limit (e.g., \citealt{10.1093/mnras/sty971}) and ultra-luminous X-ray binaries sources (e.g., \citealt{begelman2006}).\\
One class of AGN thought to be accreting at high rates are Narrow-Line Seyfert 1 (NLS1) galaxies. In fact NLS1s have a black hole mass which is typically lower than other AGN (i.e. $10^6-10^8 M_{\odot}$, \citealt{1999ApJ...521L..95P}) and since their bolometric luminosity is comparable to that of more massive AGN, they must accrete in regimes close to the Eddington limit\citep{Komossa_2006,2018rnls.confE..15K,2018rnls.confE..34G,2020Univ....6..136F}. NLS1s are characterized by relatively narrow broad emission lines, conventionally full-width at half maximum (FWHM) of H$\beta$ < 2000\,km\,s$^{-1}$, strong Fe II lines, weak [OIII] lines and steep 2--10\,keV spectra \citep{1996A&A...305...53B,Veron2001} NLS1s galaxies properties have been intensively studied over the years \citep{Pounds1995,1997MNRAS.285L..25B,2000NewAR..44..503L,Boller_2002}
but despite the recent advances in theoretical (e.g., \citealt{Jiang2019,Okuda_2021}) and observational efforts (e.g., \citealt{Wang2014a,Du2018} and references therein) not many extensive studies on the X-ray broad-band spectra of a sample of super-Eddington sources with such extreme values of accretion rates have been performed so far.\\
Here we report the X-ray spectral analysis of the joint \xmm or \textit{Swift}/XRT and \nustar observations of a sample of eight super-Eddington sources. These sources are part of a joint \xmm\ and \nustar campaign that aims to study the broad-band X-ray properties of super-Eddington AGN from the Super-Eddington Accreting Massive Black Holes sample (SEAMBHs, \citealt{Du2014,Wang2014a,2015ApJ...806...22D}) which includes objects with black hole masses estimated from reverberation mapping. For all the sources of the sample the dimensionless accretion rate $\dot{\mathscr{M}}$ can be estimated through the physics of thin accretion disks, from the part of the spectrum where $L_{\nu} \propto M^{4/3} \dot{\mathscr{M}}^{2/3} \nu^{1/3}$ regardless of the value of the BH spin (see \citealt{2002A&A...388..771C,2011ApJ...728...98D} and references therein), using the following equation: $\dot{\mathscr{M}}=20.1\,\mathscr{L}_{44}^{3/2}M_7^{-2}$ from the Shakura-Sunyaev disk model \citep{2015ApJ...806...22D}, where $\mathscr{L}_{44}$ is the 5100\AA\, luminosity in units of $10^{44}\,{\rm erg\,s^{-1}}$ and $M_7=M_{\rm BH}/10^7M_{\odot}$. This approximation is valid for $\dot{\mathscr{M}}\lesssim 10^3$ and comes from the classical Shakura \& Sunyaev thin accretion disks system. However, this can be considered valid also for slim disks. In fact the observed 5100\AA\, emission comes from large disk radii, thus this quantity it is not affected much by the radial motion. At these large radii, due to the domination of Keplerian rotation, the release of gravitational energy due to viscosity is balanced by radiation cooling, thus, all effects arising from radial advection and the black hole spin can be neglected.\\
Details about the analysed sources are reported in Table \ref{tab:sources_table}. The black hole masses of the sources analysed in this paper are estimated via reverberation mapping which is a method, suggested by \citealt{1972ApJ...171..467B}, used to map the gas distribution and derive several fundamental properties in AGN by measuring the delayed response of the broad emission line gas to the ionizing continuum. Their bolometric luminosities, \citep{Kaspi2000,Bentz2009,2016MNRAS.458.1839C}, are computed with a spectral energy distribution (SED) fitting procedure which takes into account the correction for intrinsic reddening and host-galaxy contribution.\\
\begin{table*}
\centering
\caption{Details of the sources of the sample.}
\begin{tabular}{lccccccc}
\hline\hline
Source & z & $\rm N_{\rm H}^{\rm Gal}$&$\log(M_{\rm BH}/M_{\odot})$&$\log(\dot{\mathscr{M}}$)&$\log(L_{\rm bol})[erg/s]$&$\log(\lambda_{\rm Edd})$&Ref.\\
\hline\hline
\small{IRAS\,04416+1215}&0.0889 & 12.50 &$6.78^{+0.31}_{-0.06}$&$2.63^{+0.16}_{-0.67}$&47.55&2.67&1,2,5 \\
\small{IRASF\,12397+3333} &0.0435& 1.40 &$6.79^{+0.27}_{-0.75}$& $2.26^{+0.98}_{-0.62}$&46.92&2.03&1,2,5\\
\small{Mrk\,493} &0.0313 & 1.96 & $6.14^{+0.04}_{-0.11}$& $1.88^{+0.83}_{-0.21}$&46.41&2.17&1,2,5\\
\small{Mrk\,142} &0.0449 & 1.33 & $6.59^{+0.07}_{-0.07}$&$1.65^{+0.23}_{-0.23}$&46.60&1.91&1,2,5\\
\small{Mrk\,382} & 0.0337& 4.69 & $6.50^{+0.19}_{-0.29}$&$1.18^{+0.69}_{-0.53}$&45.41&0.81&1,2,5\\ 
\small{PG\,0026+129} &0.142 & 4.73 &$8.15^{+0.09}_{-0.13}$&$0.65^{+0.28}_{-0.20}$&44.84&0.45&3 \\
\small{PG\,0953+414} & 0.234 & 1.09 &$8.44^{+0.06}_{-0.07}$&$0.39^{+0.16}_{-0.14}$&45.07&0.39&3\\
\small{NGC\,4748} & 0.0146& 3.56 &$6.61^{+0.11}_{-0.23}$&$0.10^{+0.61}_{-0.44}$&42.56&0.07&4 \\
\hline \hline
\end{tabular}\\
\label{tab:sources_table}
{\raggedright \small{\textbf{Note:} $\rm N_{\rm H}(10^{20}\rm cm^{-2})$ is the Galactic column density at the position of the source \citep{HI4PI2016}.\\ The dimensionless accretion rate \citep{Du2014}, is defined as $\dot{\mathscr{M}}\equiv\dot{M}_{\rm BH}c^2/L_{\rm Edd}$, where $\dot{M}_{\rm BH}$, is mass accretion rate, $c$ is speed of light and $L_{\rm Edd}$ is the Eddington luminosity. It is estimated by $\dot{\mathscr{M}}=20.1\,\mathscr{L}_{44}^{3/2}M_7^{-2}$ from the Shakura-Sunyaev disk model \citep{2015ApJ...806...22D}, where $\mathscr{L}_{44}$ is the 5100\AA\, luminosity in units of $10^{44}\,{\rm erg\,s^{-1}}$ and $M_7=M_{\rm BH}/10^7M_{\odot}$. This approximation is valid for $\dot{\mathscr{M}}\lesssim 10^3$.}\par}
{\raggedright \small{\textbf{References:} (1) \citealt{Du2014}, (2) \citealt{Hu2015},(3) \citealt{Kaspi2000}, (4) \citealt{Bentz2009}, (5) \citealt{2016MNRAS.458.1839C}.} \par}
\end{table*}
The paper is organized as follows. In Section \S \ref{sect:obs_data_reduction}, we present the X-rays observations and the data reduction of the sources of our sample. In Section \S \ref{sect:spectralanaysis} we describe the spectral data analysis processes and our best-fitting results. In Section \S \ref{sect:discussion} we discuss the results of our analysis which are summarized in Section \S \ref{sect:conclusion}.\\ 
Standard cosmological parameters (H=70\,km\,s$^{-1} \rm Mpc^{-1}$, $\Omega_{\Lambda}$=0.73 and $\Omega_m$=0.27) are adopted throughout the paper.\\
\section{Observations and data reduction}
\label{sect:obs_data_reduction}
IRAS\,04416+1215, IRASF\,12397+3333, Mrk\,142, Mrk\,382 and Mrk\,493 have been observed during \nustar Cycle 5 simultaneously with \xmm (P.I. C. Ricci). NGC\,4748, PG\,0026+129 and PG\,0953+414 have been observed during \nustar Cycle 6 simultaneously with \textit{Swift}/XRT target of Opportunity (TOO) observations (P.I. A. Tortosa). A brief description of the targets of this sample is reported in Appendix \ref{app:targets}.\\
The \xmm observations were performed with the European Photon Imaging Camera (EPIC hereafter, \citealt {EPIC}) detectors, and with the Reflection Grating Spectrometer (RGS hereafter; \citealt{denHerder2001}). The EPIC cameras were operated in small window and thin filter mode. \nustar telescope \citep{NuSTAR} observed all the sources with its two coaligned X-ray telescopes Focal Plane Modules A and B (FPMA and FPMB, respectively). The \textit{Swift}/XRT observations were performed as Targets of Opportunity with the Swift X-ray Telescope (XRT, \citealt{XRT}) in PC Mode, simultaneously with the \nustar observations. Details on duration and exposure of the observations are reported in Table \ref{tab:observations_table}.\\
The event lists of the EPIC cameras, both pn \citep{Struder2001} and MOS \citep{Turner2001}, are extracted with the \textsc{epproc} and \textsc{emproc} tools of the standard System Analysis Software (\textsc{SAS} v.18.0.0; \citealt{Gabriel2004}). The extraction radii and the optimal time cuts for flaring particle background were computed via an iterative process which maximizes the SNR, similar to the approach described in \citet{Piconcelli2004}. The spectra were extracted after checking that no significant pile-up affected the data, as indicated by the SAS task \textsc{epatplot}; the resulting optimal extraction radius was $30\arcsec$ and the background spectra were extracted from source-free circular regions with radii of $\sim$ $60\arcsec$. Response matrices and auxiliary response files were generated using the SAS tools \textsc{rmfgen} and \textsc{arfgen}, respectively. All the spectra were binned in order to over-sample the instrumental resolution by at least a factor of three and to have no less than 20 counts in each background-subtracted spectral channel.\\
The \nustar Level 1 data products were processed with the \nustar Data Analysis Software (\textsc{NuSTARDAS}) package (v.1.9.7) within the \textsc{heasoft} package (version 6.30). Cleaned event files (level 2 data products) were produced and calibrated using standard filtering criteria with the \textsc{nupipeline} task, and the latest calibration files available in the \nustar calibration database (CALDB 20211020). For both FPMA and FPMB the radii of the circular region used to extract source and background spectra were $40\arcsec$ and $60\arcsec$, respectively. The low-energy (0.2--5\,keV) effective area issue for FPMA \citep{Madsen2020} does not affect our observations, since no low-energy excess is found in the spectrum of this detector. Both \nustar FPMA and FPMB spectra were binned in order not to over-sample the instrumental resolution more than a factor of 2.5 and to have a SNR greater than 5$\sigma$ in each spectral channel.\\
\textit{Swift}/XRT spectra were extracted using the \textsc{xselect} line interface (v2.4k) within the \textsc{heasoft} package (version 6.28). The background extractions are measured in an annular region with radius from $30\arcsec$to $60\arcsec$. If there is pile-up, the measured rate of the source is high (above $\sim$0.6 counts\,s$^{-1}$ in the Photon-Counting Mode). The \textit{Swift}/XRT spectra of the sources of our sample resulted to not have a high pile-up degree, the sources extraction regions are measured using a circular region with a radius of $30\arcsec$.\\
\section{Spectral Analysis}
\label{sect:spectralanaysis}
We performed the spectral analysis using the version v.12.12.0 of the \textsc{xspec} software package \citep{Arnaud1996}. All errors and upper/lower limits are calculated using $\Delta\chi^2$ = 2.71 criterion (corresponding to the 90\% confidence level for one interesting parameter), and the Solar abundances of elements are assumed, if not stated otherwise. A cross-normalization between the EPIC and FPMA/B spectra is always included to account also for the slightly different average flux. The two \nustar modules (FPMA and FPMB) spectra are fitted simultaneously, with a cross-normalization constant typically less than 5\% \citep{Madsen2015}. In all the fits we included always the Galactic column density at the position of the sources, $N_{\rm H}^{\rm Gal}$, and it is modeled with the \textsc{Tbabs} component \citep{Wilms2000} with $N_{\rm H}$ frozen to the nominal value (see second column of Table \ref{tab:sources_table} for more details).\\
Our spectral analysis is focused on the 0.3--3\,keV range for \textit{Swift}/XRT spectra, on the 0.3--10\,keV range for \xmm spectra and on the 3--25\,keV range for the \nustar spectra since the spectra are background dominated for energies outside these ranges. All the spectra we used for this analysis have been corrected for the effective area of each detector and have been re-binned to have at least 20 counts per bin to use chi-squared statistics. 
\subsection{General description of the fitting models}
\label{sect:fitting_models}
\begin{table}
\centering
\caption{List of the models used in the fitting analysis.}
\begin{tabular}{ll}
\hline\hline
Model name &  \textsc{xspec} model composition\\
\hline
A & zTbabs * WAs * (bbody+cut-offpl+pexrav+zgauss)\\
B1 &  zTbabs * WAs * (bbody+xillver)\\
B2 &zTbabs * WAs * (bbody+xillverD)\\
B3 & zTbabs * WAs * (bbody+xillverCp)\\
C1 & zTbabs * WAs * (bbody+relxill)\\
C1 + NL & zTbabs * WAs * (bbody+relxill+zgauss)\\
C2 & zTbabs * WAs * (bbody+relxillD)\\
C3 & zTbabs * WAs * (bbody+relxillCp)\\
C4 & zTbabs * WAs * (relxill)\\
D & zTbabs * WAs * (bbody+relxill+xillver)\\
\hline\hline
\end{tabular}\\
\label{tab:models}
\end{table}
\begin{table*}
\centering
\caption{Best-fitting parameters for the X-ray broad-band \textit{XMM-Newton} plus \textit{NuSTAR} spectra of our sample of super-Eddington sources for the primary continuum and the reprocessed emission. Errors are at 90\% confidence levels. If a model is calculated by default for a fixed value of a parameter, that value is shown in boldface.}
\begin{tabular}{lcccccccc}
\hline\hline
\noalign{\smallskip}
Source & Model$^{\star}$ & $\Gamma$ & $E_{\mathrm{cut}}$(keV) & $R_{\rm refl}$ & \logxi & A$_{\rm Fe}$ & $\mathit{a}$ & $\chi^2$/dof\\
\hline\hline
\noalign{\smallskip}
IRAS\,04416+1215 & B1 &$1.77_{-0.09}^{+0.17}$& $44_{-17}^{+28}$ & $>8.01$ & $3.97_{-0.30}^{+0.25}$ & $5.01_{-2.63}^{+2.98}$ & \textbf{0} & 1.08\\
\noalign{\smallskip}
IRASF\,12397+3333& C1 &$2.28\pm0.02$ & NC & $0.75^{+0.52}_{-0.38} $ & $3.30^{+0.07}_{-0.10}$ & $0.93^{+0.21}_{-0.14}$ & $0.78^{+0.18}_{-0.29}$ & 1.05 \\
\noalign{\smallskip}
Mrk\,493& C1 & $2.08\pm 0.04$ & NC & $1.12 \pm 0.14$ & $3.86_{-0.06}^{+0.07}$ & $3.86_{-0.31}^{+0.34}$ & $<0.80$ & 1.05\\
\noalign{\smallskip}
Mrk\,142& C1 & $2.37\pm0.05$ & NC & $1.18\pm0.16$ & $3.03\pm0.06$ & $>9.40$ & $0.1\pm0.59$ & 1.09\\
\noalign{\smallskip}
Mrk\,382& C1 &$2.02\pm0.03$ & $45^{+7}_{-8}$ & $3.52^{+0.54}_{-0.76} $ & $1.28\pm0.15$ & $2.98\pm0.77$ & $>0.89$&1.02\\
\noalign{\smallskip}
PG\,0026+129 & C1 & $2.10 \pm 0.05$ & NC & $1.45\pm0.35$ & $2.50\pm0.19$ & $<0.53$ & $>0.65$ &1.02\\
\noalign{\smallskip}
PG\,0953+414& B1 & $2.41\pm0.02$ & NC & $2.44\pm0.18$ & $0.02^{+0.03}_{-0.11}$&$<0.52$&\textbf{0}& 1.09\\
\noalign{\smallskip}
NGC\,4748& B1 & $1.98\pm0.04$ & $>158$ & $0.47^{+0.17}_{-0.09}$& $1.20\pm0.96$& $1.87^{+1.12}_{-1.27}$ & \textbf{0}& 1.01\\
\noalign{\smallskip}
\hline \hline
\end{tabular}\\
\label{tab:best-fitting-param}
\smallskip
{\raggedright \footnotesize{($\star$) The fitting models are described in Section \ref{sect:fitting_models} and listed in Table \ref{tab:models}.} \par}
\end{table*}
We performed the X-ray broad-band spectral fitting following the same steps outlined in Section 4 of \citet{dinosaur}. The soft energy band (i.e., 0.3--2\,keV) has been analysed including \textit{XMM-Newton} EPIC-pn and MOS 1+2 spectra simultaneously to better constrain the parameters of the warm absorbing and soft-excess components. We tied all the MOS parameters to the pn values, apart from the normalization of the various components.
Once good constraints are obtained for the soft energy band components we excluded the MOS 1+2 data, since they do not provide any significant improvement to the statistic for constraining the higher energy band components.\\
All the sources of our sample show the presence of a soft excess component and warm absorbing components. To take into account the absorption components we included in the fitting model at first one (or more than one depending on the case) \textsc{zxipcf} model for partial covering of partially ionized absorbing material. This model uses a grid of \textsc{xstar} \citep{Kallman2001} photoionized absorption models (calculated assuming a turbulent velocity of 200\,$\rm km\,s^{-1}$) for the absorption, assuming that the absorber only covers some fraction of the source. For a more refined modeling of these absorbers we replaced the \textsc{zxipcf} components with detailed detailed grids computed with the photoionization code \textsc{xstar}, with a spectral energy distribution described by a power-law with a photon index of $\Gamma = 2$ and turbulent velocity of 100\,$\rm km\,s^{-1}$ and/or 1000\,$\rm km\,s^{-1}$. These tables consider standard solar abundances from \citet{Asplund2009}, and take into account absorption lines and edges for all the metals characterized by an atomic number $Z \leq 30$. Regardless of the number of absorption components and for sake of simplicity, we will refer to the warm absorbing components as WAs throughout all the paper. All the spectral component are taken into account in the best fitting model but the detailed analysis and spectral parameters values of the warm-absorption and outflows components will be presented in a forthcoming paper (Tortosa et al., in prep) while here we will focus mostly on the analysis of the primary continuum and the reflection component.\\
In the broad-band analysis we did not included the MOS 1+2 spectra to simplify the analysis since in this work we are mostly interested to the primary continuum and reflection properties.\\
The fit procedure was carried out testing different models (see Table \ref{tab:models}) for the primary continuum and the reprocessed radiation. The first model we tested was a simple phenomenological model (Model A) in which the primary continuum and the reprocessed emission are modelled with the \textsc{cutofpl} (power-law with an exponential cut-off at high energy) and \textsc{pexrav} \citep{pexrav} models in \textsc{xspec} respectively. In this model we kept the abundances of the elements fixed to Solar values and the inclination angle fixed to 30$^{\circ}$. We also tested the presence of the Iron K$\alpha$ emission line by including a simple Gaussian line profile in the model (\textsc{zgauss} model in \textsc{xspec}). In this test the energy of the line as well as the line width were free to vary.\\
After testing this simple phenomenological model, we modeled the primary continuum and the reprocessed emission testing the standard reflection using the photoionized reflection model \textsc{xillver} version [1.4.3] \citep{Garcia2013}, which accounts also for the Fe K$\alpha$ emission line and the relativistic reflection model \textsc{relxill} v.1.4.3 \citep{Garcia2014, Dauser2014}. We tested the different flavours of the \textsc{xillver} and \textsc{relxill} models. First we tested the standard (Model B1) and relativistic (Model C1) reflection models, then the standard (Model B2) and relativistic (Model C2) reflection models but allowing a higher density for the accretion disk, between $10^{15}$ to 10$^{19}$ cm$^{-3}$, using respectively \textsc{xillverD} and \textsc{relxillD} models. We checked for the presence of a narrow component of the Iron K$\alpha$ emission line by including a simple Gaussian line profile in the Model C1 (Model C1+NL). We extrapolated the equivalent width of the narrow component of the Iron K$\alpha$ in this model to look for correlation of this quantity with the Eddington ratio and with the 2--10\,keV luminosity. Then, to obtain a direct measurement of the coronal temperature, we tested standard (Model B3) and relativistic (Model C3) reflection models given for an incident spectra with a \textsc{nthcomp} Comptonization model \citep{Zdziarski1996, Zycki1999}, using \textsc{xillverCp} and \textsc{relxillCp} models respectively. In both the standard and relativistic reflection models \textsc{xillverD} and \textsc{relxillD} the high-energy cut-off is fixed at 300 keV by default. In all the different flavours of \textsc{xillver} and \textsc{relxill} the disk inclination was fixed to 30$^{\circ}$ while Fe abundance, ionization parameter and reflection fraction were allowed to vary. In the models including the \textsc{relxill} component the BH spin parameter was allowed to vary while both emissivity indices $q_1$ and $q_2$ were fixed to the non relativistic value of $q_1=q_2=3$. The disk inner radius was set to $R_{\rm in,disk}\equiv \rm R_{\rm isco}$. Since the basic \textsc{relxill} model does not account for the soft X-ray excess we tested also the \textsc{relxill} model allowing for a broken power-law emissivity function (Model C4). In this model the break radius $R_{\rm br}$ and the inner emissivity index q$_1$ are free to vary, while the outer emissivity index is still fixed to the standard limit of q$_2$=3. Since this model accounts for the soft excess component we did not include the black-body component.\\
We also checked for the presence of a distant reflection component (Model D). This can be done by including a blurred reflection component for the disk (\textsc{relxill}) and an unblurred neutral reflection component (\textsc{xillver}). The photon indices and the cut-off energies of the two components are tied together. Solar iron abundance and \logxi=0 are assumed for the latter. 
\subsection{Broad-Band Analysis Results}
\label{sect:best_fit}
\begin{figure*}
	\includegraphics[scale=0.3]{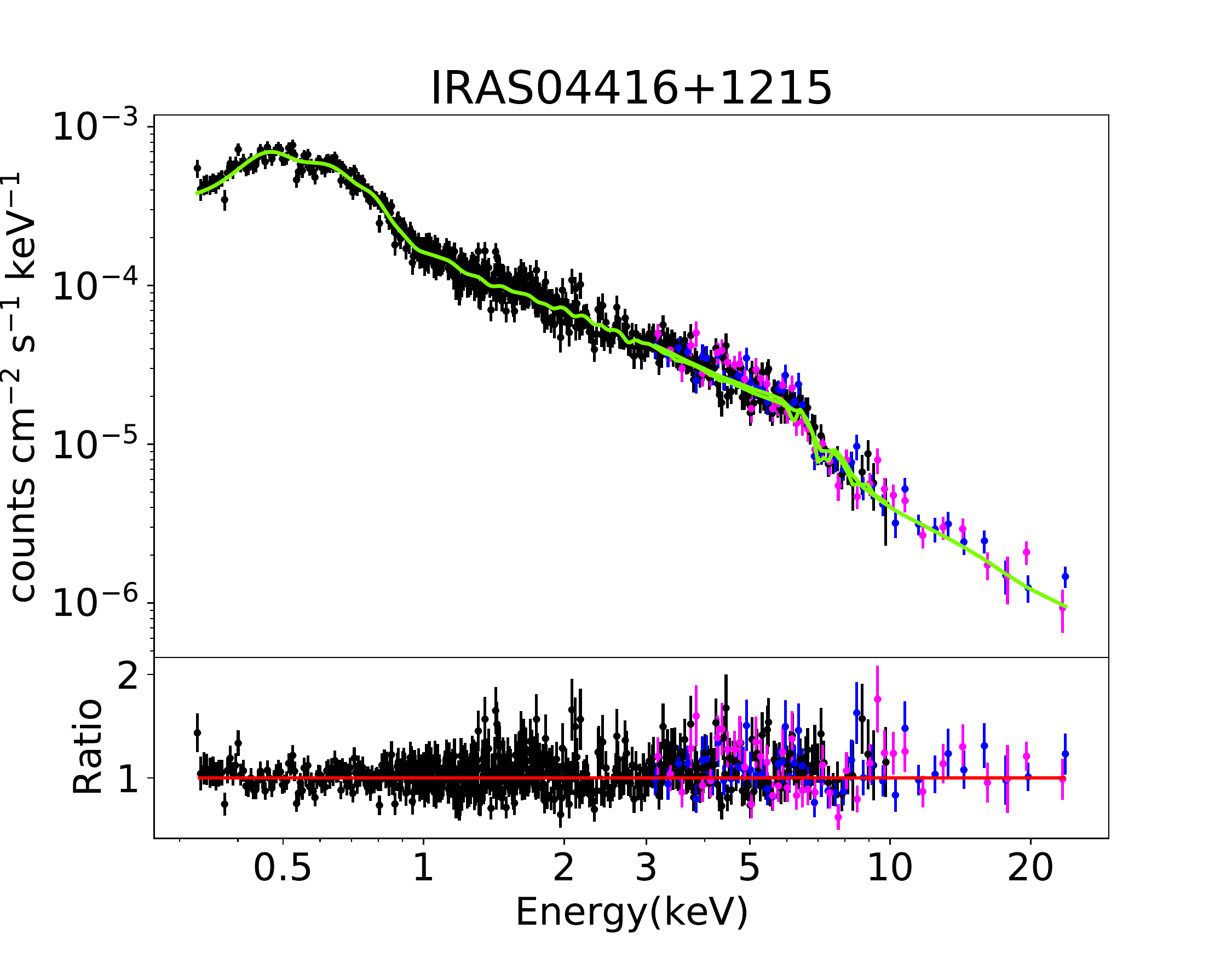}
	\includegraphics[scale=0.3]{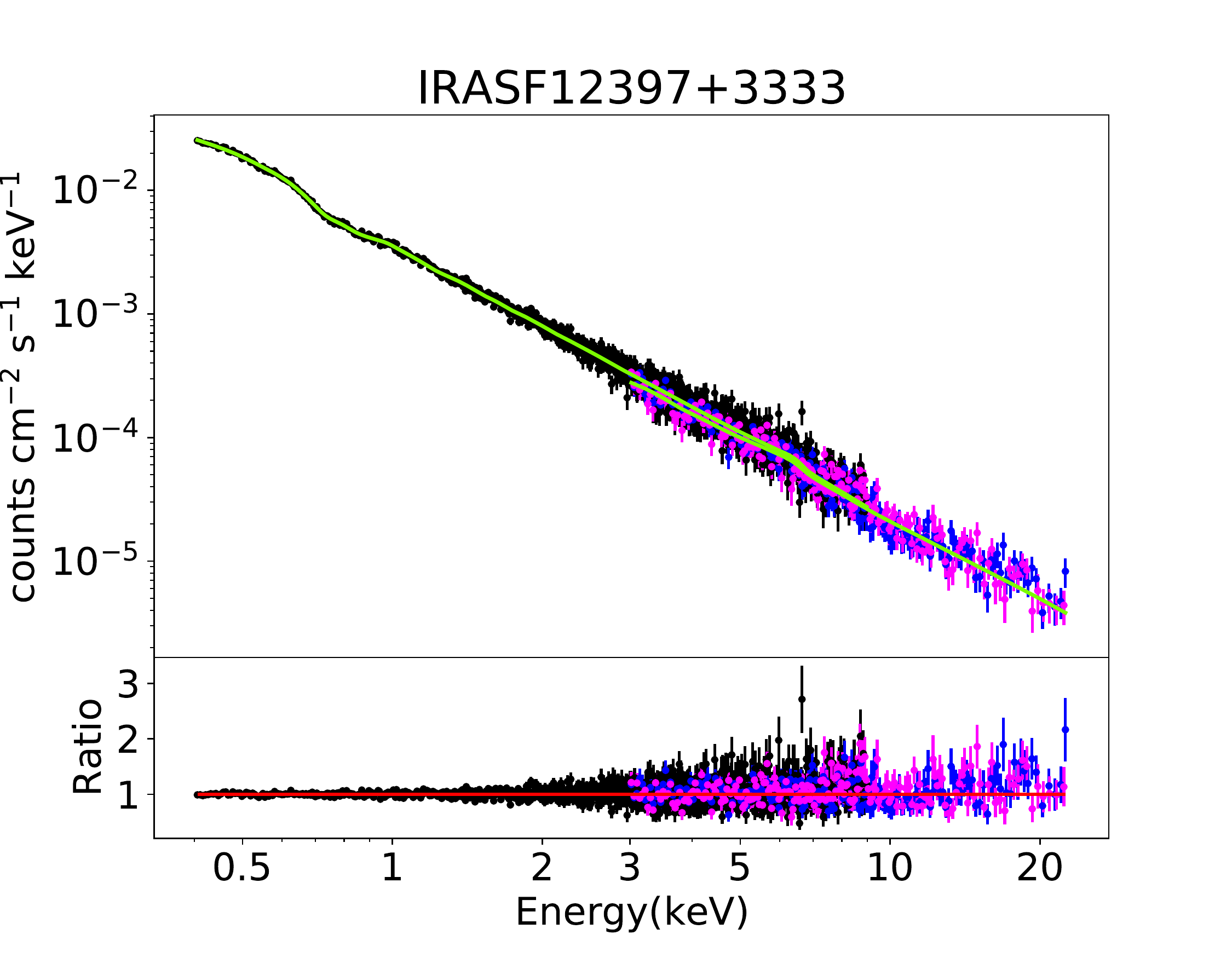}
	\includegraphics[scale=0.3]{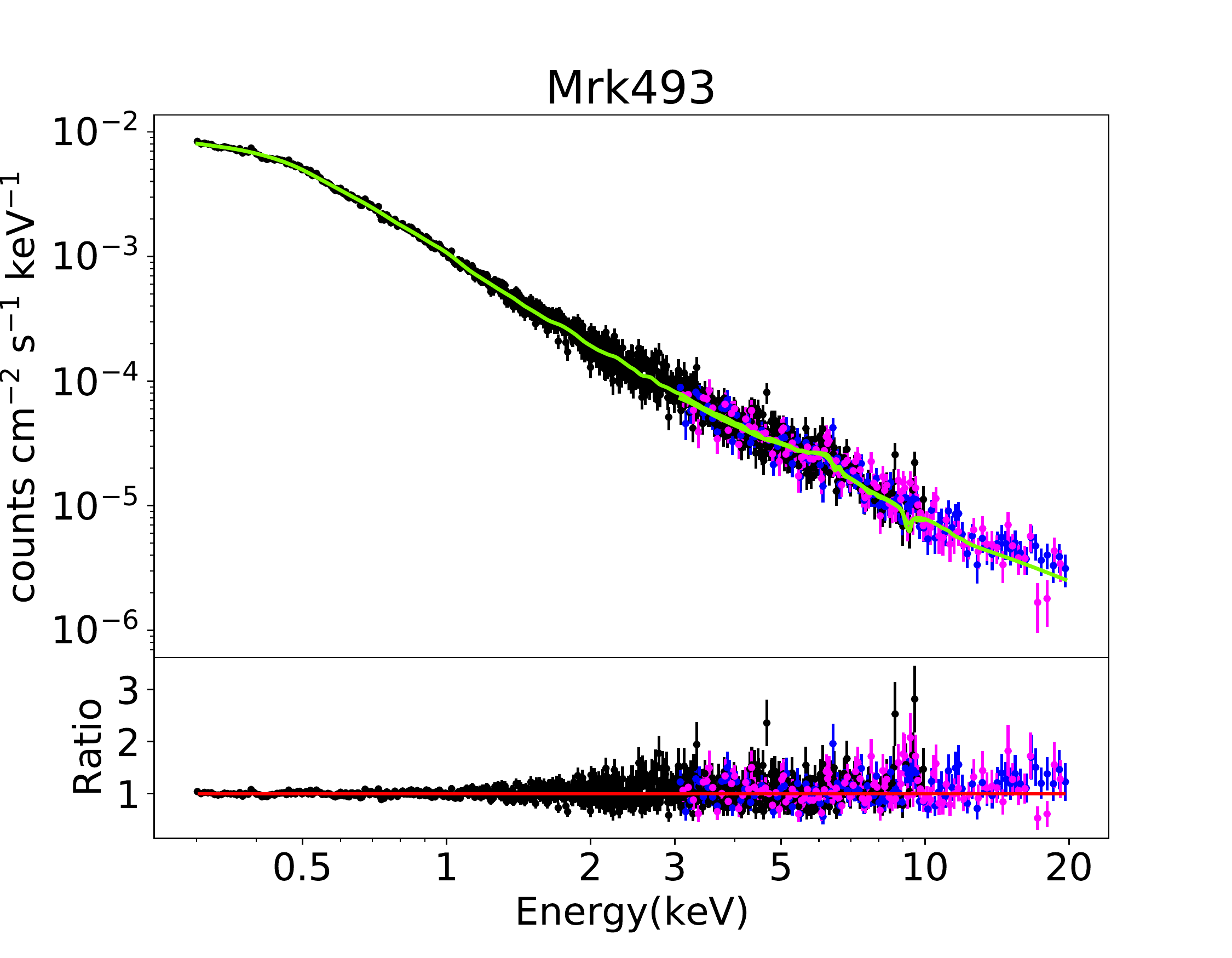}
	\includegraphics[scale=0.3]{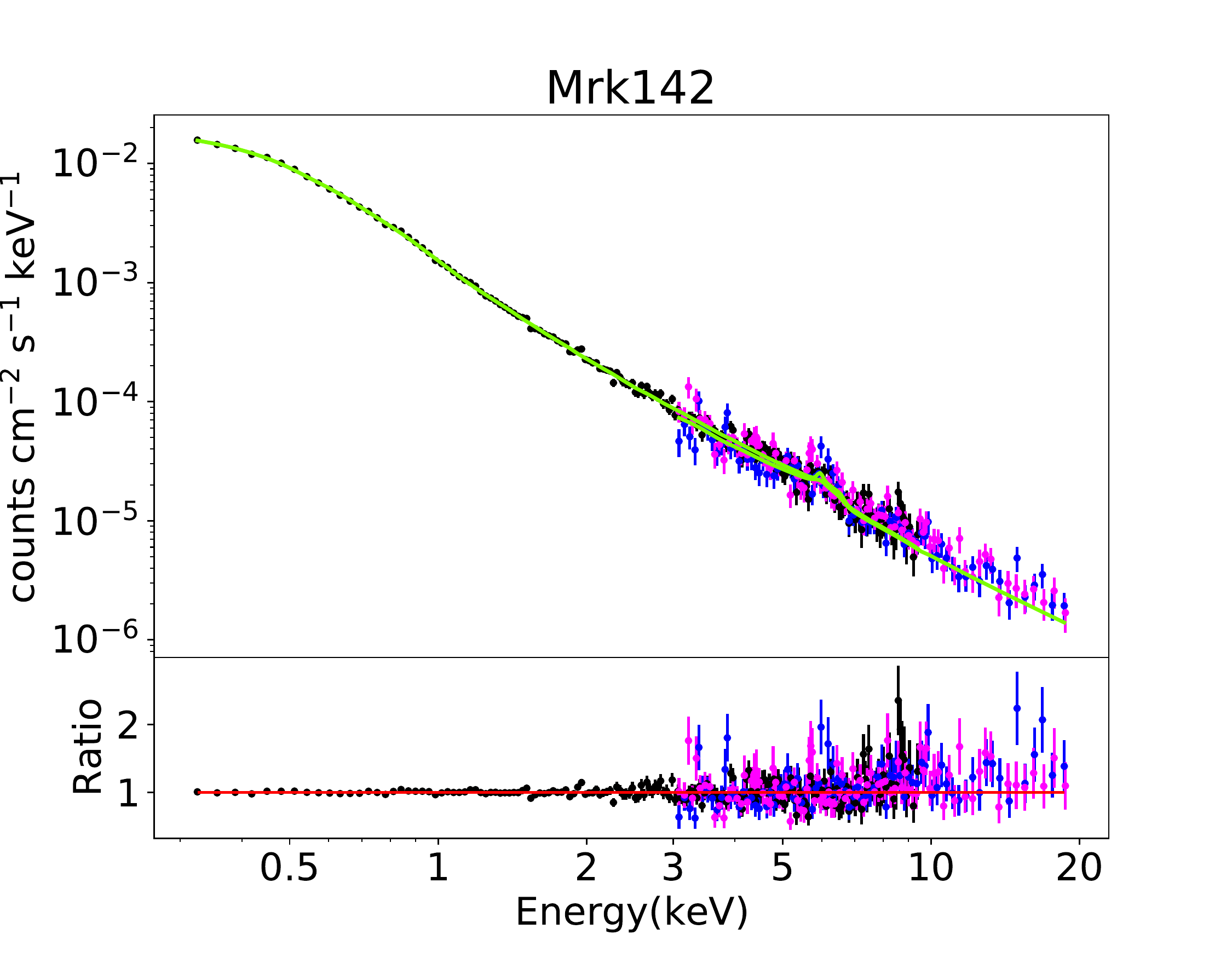}
	\includegraphics[scale=0.3]{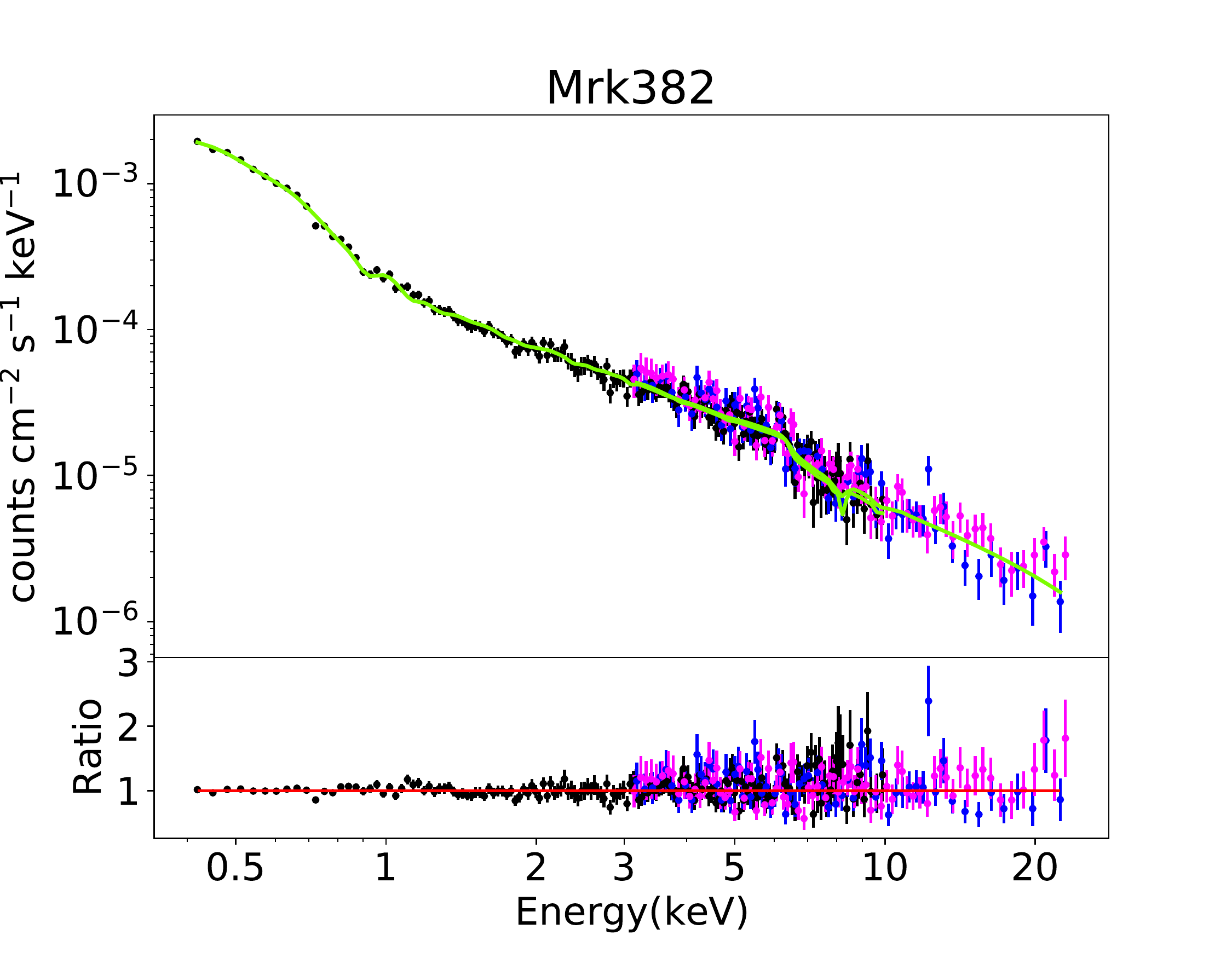}
	\includegraphics[scale=0.3]{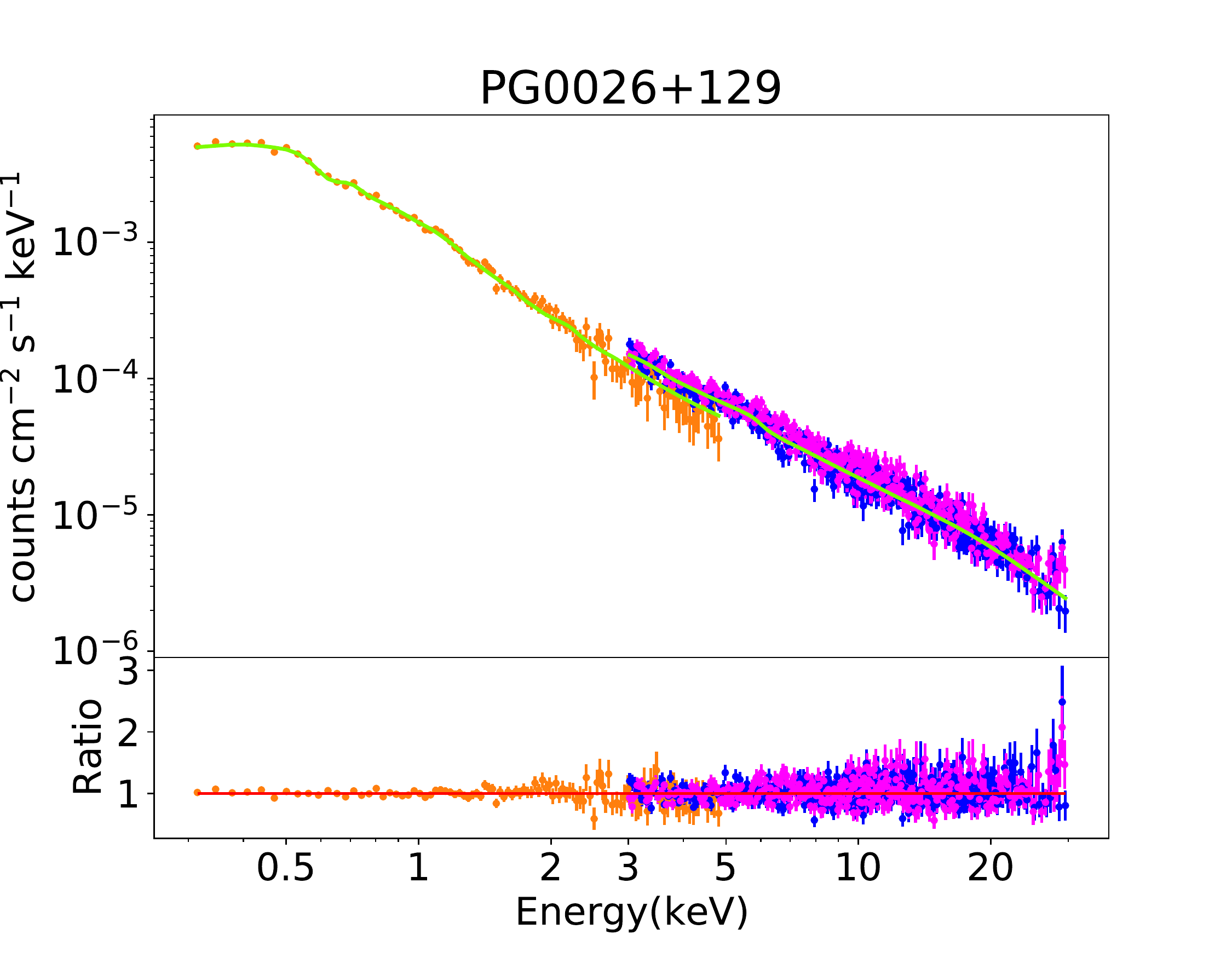}
	\includegraphics[scale=0.3]{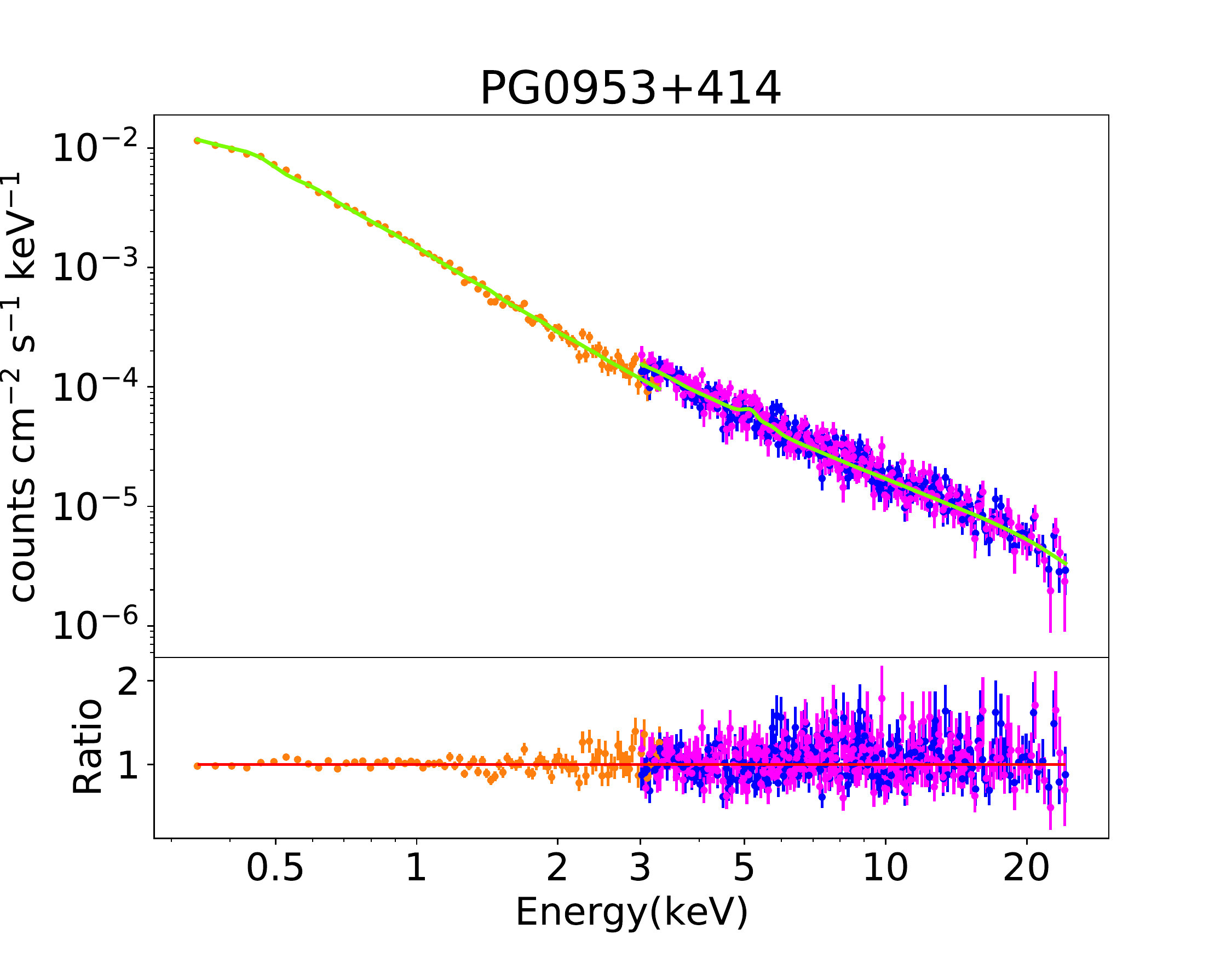}
	\includegraphics[scale=0.3]{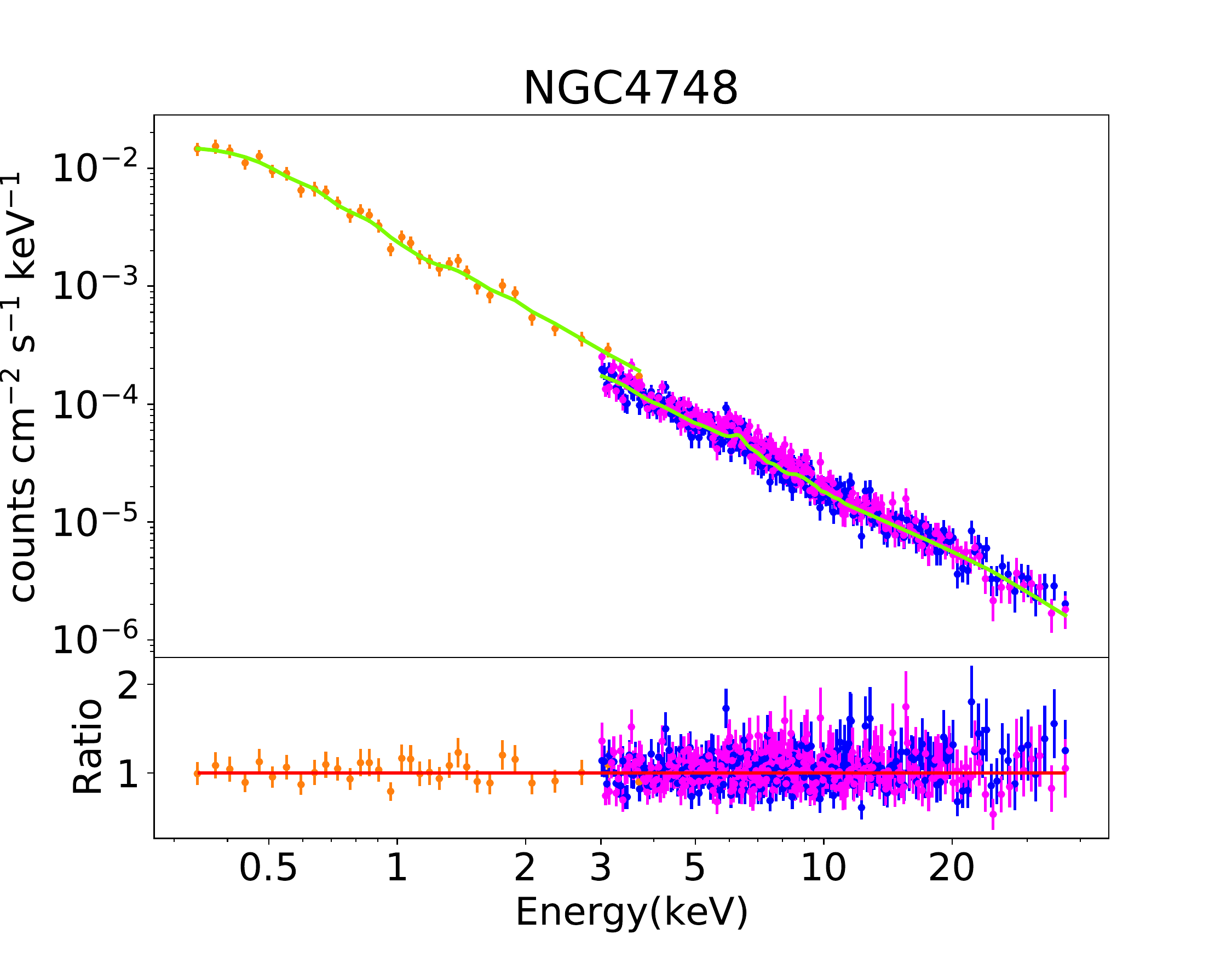}
    \caption{Spectra, fitting model (top panel) and residuals ratio (bottom panel) for EPIC-pn (black), \textit{Swift}-XRT (orange) and \textit{NuSTAR} FPMA (blue) and FPMB (magenta) spectra of the super-Eddington sources of our sample considering the best-fitting model, see Table \ref{tab:best-fitting-param}. From top left to bottom right the sources are: IRAS\,04416+1216, IRASF\,12397+3333, Mrk\,493, Mrk\,142, Mrk\,382 and  PG\,0026+129. All the spectra are background-subtracted and corrected for the effective area of each detector.}
    \label{fig:data}
\end{figure*}
The detailed analysis of the sources of our sample with all the fitting models described in previous Section \ref{sect:fitting_models} is reported in Appendix \ref{app:sources_details}. Here we report only the best-fitting results. For most of the sources of our sample the best fit model is the Model C1, except for IRAS\,04416+1215 (see \citealt{dinosaur}), PG\,0953+414 and NGC\,4748 for which the best fitting model is Model B1. The best fitting parameter values are reported in Table \ref{tab:best-fitting-param} while the plot of data, fitting models and residuals ratios are reported in Figure \ref{fig:data} . In some cases the statistical significance of the fits is the same while fitting with the Model C2, C4 or D but either or the characterizing parameters are unconstrained or the distant neutral component is unconstrained. Thus, we do not consider them as best-fitting models.\\
We searched for possible degeneracies among the fitting parameters, performing Monte Carlo Markov Chain (MCMC) using \textsc{xspec-emcee} code by Jeremy Sanders\footnote{\url{https://github.com/jeremysanders/xspec_emcee}}. This is an implementation of the emcee code \citep{ForemanMackey2013}, to analyze X-ray spectra in \textsc{xspec}. We used 50 walkers with 10,000 iterations each, burning the first 1,000. The walkers started at the best fit values found in \textsc{xspec} (see Table \ref{tab:best-fitting-param}), following a Gaussian distribution in each parameter, with the standard deviation set to the delta value of that parameter. The MCMC results of the fitting models applied to the broad-band data are shown in Figure \ref{fig:triangular} for the most relevant parameters: photon index ($\Gamma$), iron abundance ($A_{\rm Fe}$), cut-off energy ($E_{\mathrm{cut}}/\rm keV$), ionization parameter [$\log(\xi /\rm erg\,\rm cm\,\rm s^{-1}$)], reflection fraction ($R_{\rm refl}$) and black hole spin ($\mathit{a}$).\\
For the majority of the sources of this sample we do not constrain the high energy cut-off parameter, for which we found lower limit which also appears to be very high (e.g. $E_c>650$\,keV). When fixing the cut-off parameter at a most reasonable value (e.g. $E_c=200$\,keV) the statistical goodness of the fits did not change. Being the \nustar spectra truncated at 20\,keV due to the high background above this energy, we are not able to constrain this parameter, except for Mrk\,382, which show a cut-off energy at $\sim45$\,keV, similarly to IRAS\,04416+1215.\\
All the sources show a steep spectrum with the photon index of the primary power-law being $\gtrsim 2$ consistent with previous studies of super-Eddington sources  \citep{1997MNRAS.285L..25B,Brightman2013,dinosaur}. Another common feature among the sources of the sample is that most of them show an iron overabundance, except PG\,0953+414 and PG\,0026+129 which, however, are the sources with the highest SMBH mass of our sample, and show a high reflection fraction and a weak iron K$\alpha$ emission line. This result confirms what previously found in the analysis of IRAS\,04416+1215, i.e. that these super-Eddington sources may show a reflection-dominated spectrum (as in the case of the NLS1 galaxy 1H\,0707-495, see \citealt{2009Natur.459..540F} and of the Seyfert 2 galaxy IRAS\,00521-7054, see \citealt{2014ApJ...795..147R}). In fact, when the accretion rate is close or above the Eddington limit the accretion disk can be radiation-pressure dominated (e.g., \citealt{1974ApJ...187L...1L,1998ApJ...498L..13K,2001xeab.confE..71T}). When this happen the accretion disk can be clumpy instead of being flat, thus the reflection component can be be stronger \citep{Fabian_2002}.
\section{Discussion}
\label{sect:discussion}
\subsection{Reflection Features and Iron line}
\label{sect:reflection}
\begin{figure*}
	\includegraphics[width=0.6\textwidth]{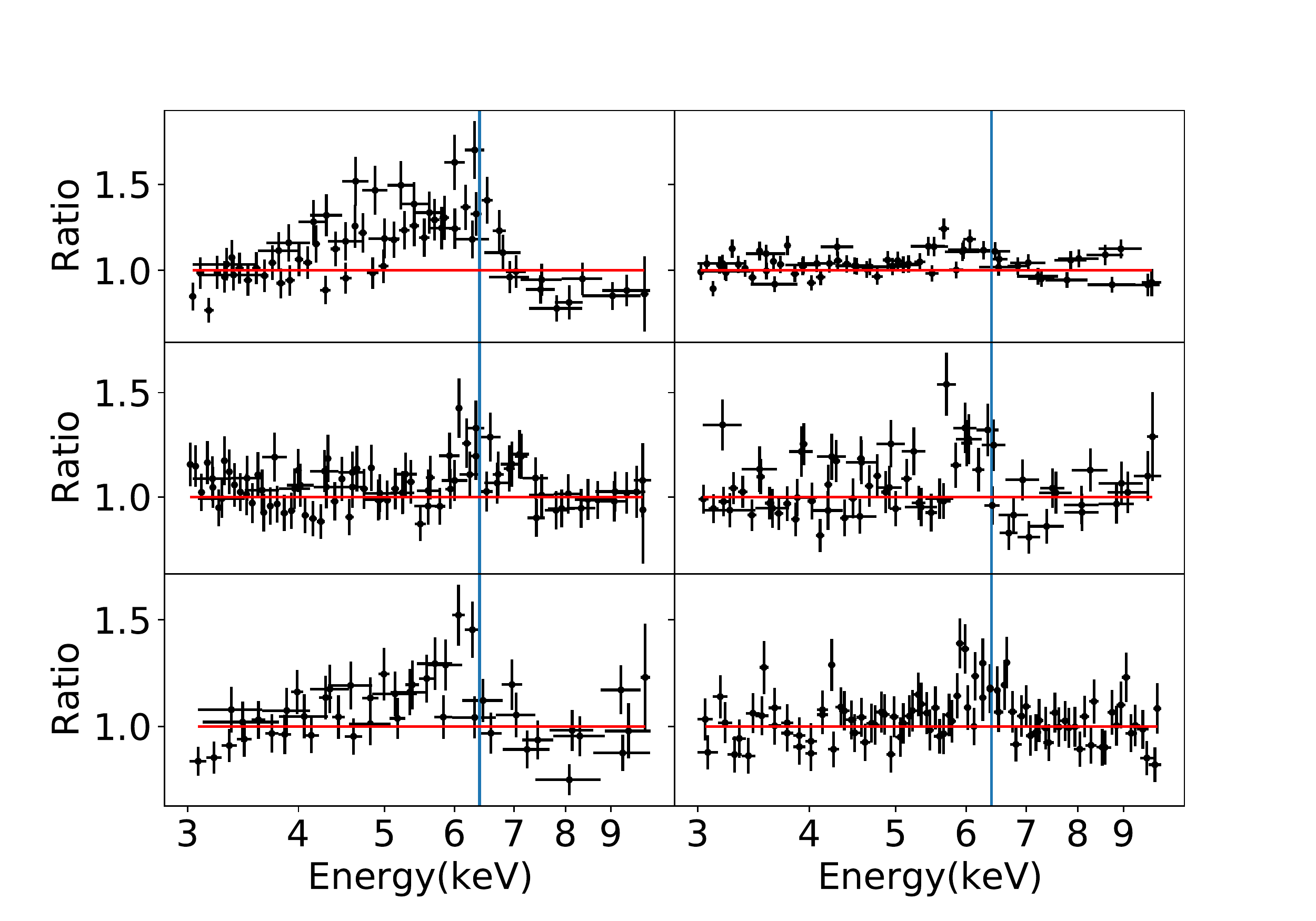}
    \caption{Ratio residuals for the EPIC-pn, FPMA and FPMB  spectra 3--10\,keV spectra when using a fitting model consisting in just a power-law for IRAS\,04416+1215 (top left panel), IRASF\,12397+3333 (top right panel), Mrk\,493 (center left panel), Mrk\,142 (center right panel), Mrk\,382 (bottom left panel), NGC\,4748 (bottom right panel). }
    \label{fig:res_iron_line}
\end{figure*}
\begin{figure*}
	\includegraphics[width=0.7\textwidth]{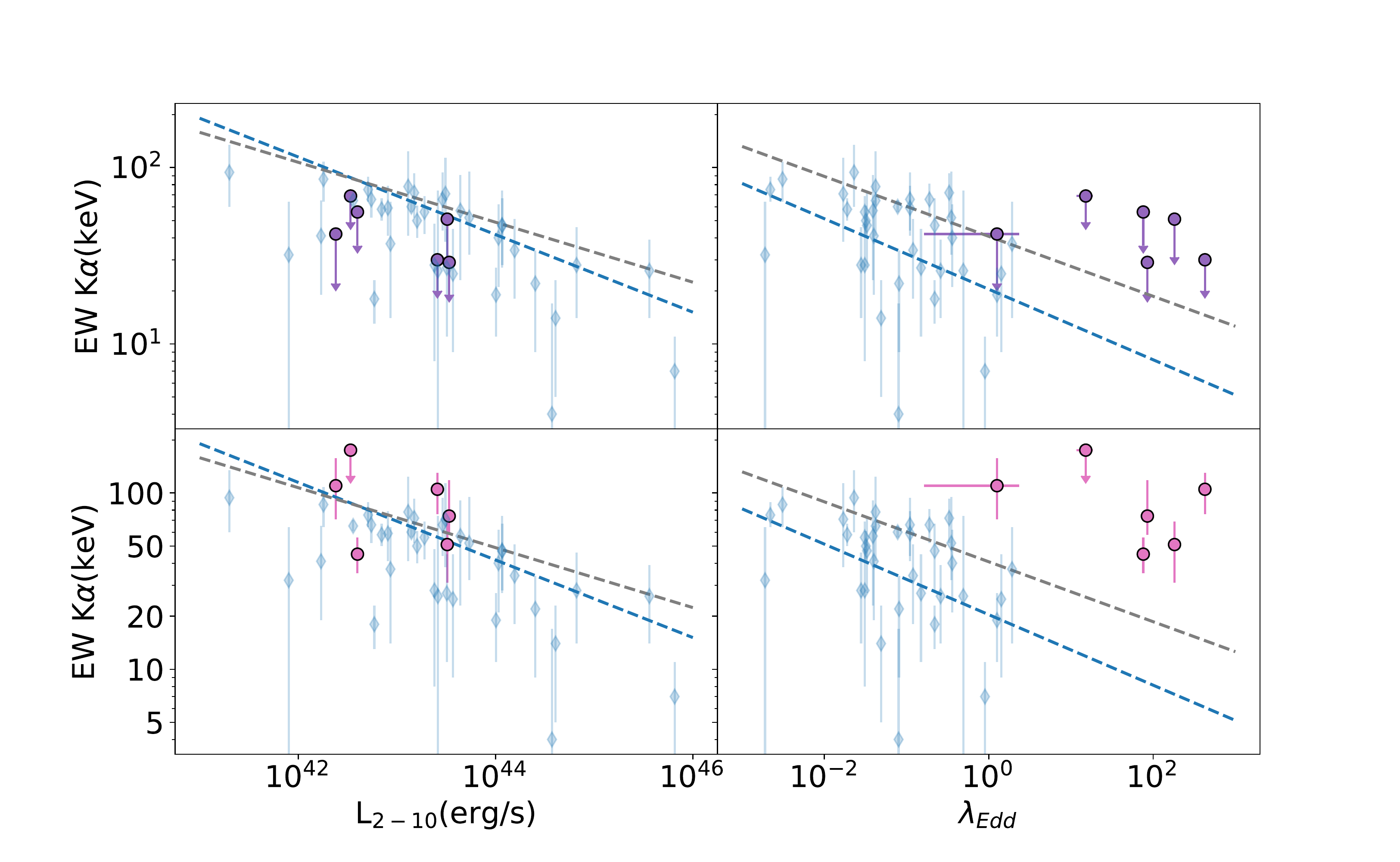}
    \caption{The Equivalent width of the neutral iron line plotted against 2--10\,keV X-ray luminosity (left panels) and versus the Eddington ratio (right panels) of the objects in our sample. In the top panels we used the EW values obtained from the fit with the best fitting Model (C1+NL) which includes also the relativistic broadening of the line. In the bottom panels we used the EW values extrapolated from the fit with a simple model consisting in a power-law plus a narrow Gaussian line (PW+GZ, see table \ref{tab:EW}). In each panel the dashed line represent the relations found by \citealt{Shu2010} (blue line), \citealt{Bianchi2007} (gray line) and the blue diamonds represent the sample from \citealt{Shu2010}.}
    \label{fig:iron_line}
\end{figure*}
The equivalent width (EW) of the Fe K$\alpha$ emission line at 6.4\,keV, is found to be inversely correlated with the X-ray luminosity in the 2--10\,keV band. This trend is known as the ‘X-ray Baldwin’ or ‘Iwasawa-Taniguchi’ effect (hereafter ‘IT effect’)  \citep{1993ApJ...413L..15I}. This inverse correlation is similar to the classical 'Baldwin effect', in which the equivalent width of the [\textsc{C IV}] $\lambda$1550 emission line is inversely correlated with the UV continuum luminosity \citep{1977ApJ...214..679B}. 
The IT effect has been found in a large sample of different objects studied with different instruments \citep{2004MNRAS.347..316P,2006ApJ...644..725J,Bianchi2007,Shu2010,Ricci2014} but all the sources belonging to these samples have $\lambda_{\rm Edd}\lesssim 2.00$. So we checked for the presence of IT effect in the super-Eddington sources of our sample by comparing the trend of the EW of the FeK$\alpha$ line versus 2--10\,keV luminosity and Eddington ratio with the results from \citet{Bianchi2007} and \citet{Shu2010}.\\
There are several explanations to the IT effect: the decrease in covering factor of the material forming the FeK$\alpha$ line \citep{2004MNRAS.347..316P}, the dependency from the luminosity of the ionisation state of the material which produce the line \citep{1997ApJ...488L..91N,2000ApJ...534..718N}, the variability related to the non-simultaneous reaction to flux changes of the continuum of the reprocessing material \citep{2006ApJ...644..725J} and possibly also to the variability of the relativistically broadened component of the iron line \citep{1997ApJ...488L..91N}. Moreover, \citet{Bianchi2007} show that the EW of the Fe K$\alpha$ line decreases with the Eddington ratio with a slope similar to that obtained using the X-ray luminosity, suggesting that possibly the IT effect can be related to the relation between the photon index of the primary power-law spectrum and the Eddington ratio \citep{2013MNRAS.435.1840R} since for a higher values of the Eddington ratio the photon index is steeper, implying the presence of less photons at the energy of the iron K$\alpha$ line and then a lower values of its EW.\\
\begin{table}
\centering
\caption{Equivalent Width values for the sources of our sample.}
\begin{tabular}{lcc}
\hline\hline
\noalign{\smallskip}
Source & EW [eV] C1+NL & EW [eV] PW+ZG\\
\hline\hline
\noalign{\smallskip}
IRAS\,04426+1215 & $<30$&$105^{+29}_{-25}$\\
IRASF\,12397+3333 &$<51$&$51^{+20}_{-18}$\\
Mrk\,493 &$<56$&$45^{+10}_{-11}$\\
Mrk\,142 &$<29$&$74^{+16}_{-44}$\\
Mrk\,382 &   $<69$&$<175$\\
PG\,0026+129& - & -\\
PG\,0953+414 & - &-\\
NGC\,4748  &$<42$&$110^{+39}_{-48}$\\
\hline \hline
\end{tabular}\\
\label{tab:EW}
\end{table}
We extrapolated the EW both from the fit of the spectra from 3 to 10\,keV using a simple power-law model which included a Gaussian line (PW+ZG) to reproduce the Iron K$\alpha$ line and from the fit of the broad-band X-ray spectra  with the Model C1+NL in the same energy band. The values are reported in Table \ref{tab:EW}. The values obtained from the fit with the simple PW+ZG Model most likely contain contamination from the broad relativistic lines produced in the inner part of the accretion flow since most of the sources of the sample show relativistic reflection features (see Figure \ref{fig:res_iron_line}). In fact we found a measure for the BH spin parameter for IRASF\,12397+3333 and for Mrk\,142 of $0.78^{+0.18}_{-0.29}$ and $0.1\pm0.59$ respectively, a lower limit of $>0.89$ and $>0.65$ for Mrk\,382 and an upper limit of $<0.80$ for Mrk\,493 (see Table \ref{tab:best-fitting-param}). For PG\,0026+129 and PG\,0953+414 we do not detected the Iron K$\alpha$ line, probably because the flux of the line was too low to be detected just with the \textit{Swift}/XRT. Thus, we did not included these sources in this analysis.\\
Considering the values of EW obtained from the PW+ZG Model, we found that the sources of our sample also show evidence of IT effect, in fact looking at the relation between the EW of the FeK$\alpha$ line versus 2--10\,keV luminosity (see lower left panel of Figure \ref{fig:iron_line}), they follow the trend of \citet{Bianchi2007} and \citet{Shu2010}. From the analysis of the relation between the EW of the K$\alpha$ line versus the Eddington ratio (see lower right panel of Figure \ref{fig:iron_line}) we found that they also follow the prediction of the literature anticorrelation. But the EW obtained by fitting the 3--25\,keV spectra with the PW+ZG Model are systematically higher with respect of what expected from the literature. This can be the effect of the enhancement of the narrow core of the Iron K$\alpha$ line due to the broad relativistic features that are not taken into account in the fit with PW+ZG Model which includes just a power-law and a narrow Gaussian line.\\
From the fit with the best fitting model (C1+NL) we found only upper limits (see upper panels of Figure \ref{fig:iron_line}) to the EW of the Fe K$\alpha$ line, thus it is not straightforward to assess that these sources follow exactly the literature predictions of the IT effect. Even if we found an iron overabundance we do not found an intense Fe K$\alpha$ line emission. Moreover, some of the sources of our sample show the presence of a sharp drop around $\sim$7\,keV, which is a typical feature of NLS1 galaxies \citep{Boller_2002}. These characteristics, together with the moderate or high values of the reflection fraction we get from the broad-band fitting analysis, suggest that the X-ray spectra of these sources are likely reflection dominated \citep{Fabian_2002}.
In a situation of extremely rapid accretion we expect a reflection dominated spectrum. In fact when the accretion rate is close or above the Eddington limit the accretion disk is most likely radiation-pressure dominated (e.g., \citealt{1974ApJ...187L...1L,1998ApJ...498L..13K,2001xeab.confE..71T}). In this scenario there can be disk instabilities, the disk could be clumpy and the reflection component could be stronger \citep{Fabian_2002}. Moreover, the large disk scale height expected in Super-Eddington AGN could also contribute to enhancing the amount of reprocessed X-ray radiation. Indeed, the median value of the reflection fraction of the SEAMBHs sources of our sample is a factor $\sim 2$ higher compared to the median value of the reflection fraction of the Sy1 sources from the BASS sample (see bottom left panel of Figure \ref{fig:primary_cont_param}).
\subsection{Primary Continuum Properties}
\label{sect:continuum}
\begin{figure*}
	\includegraphics[width=1\textwidth]{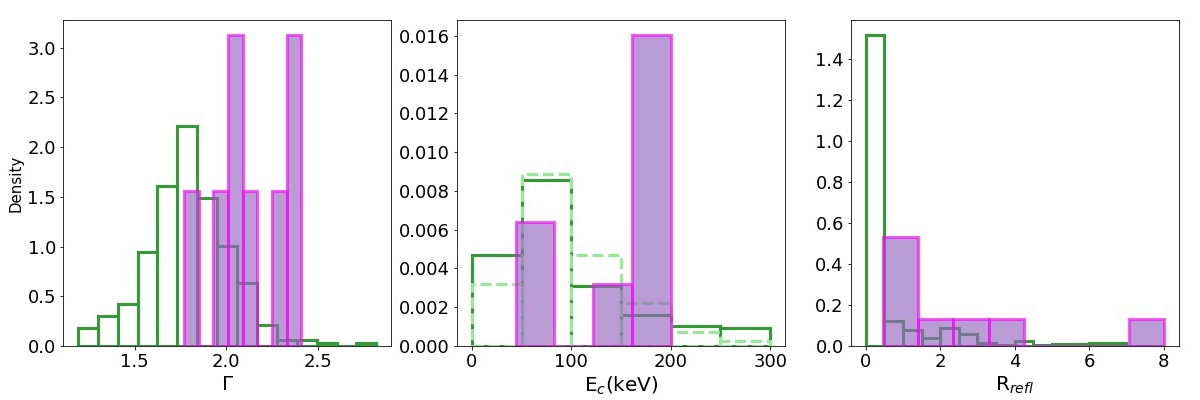}
	\includegraphics[width=1\textwidth]{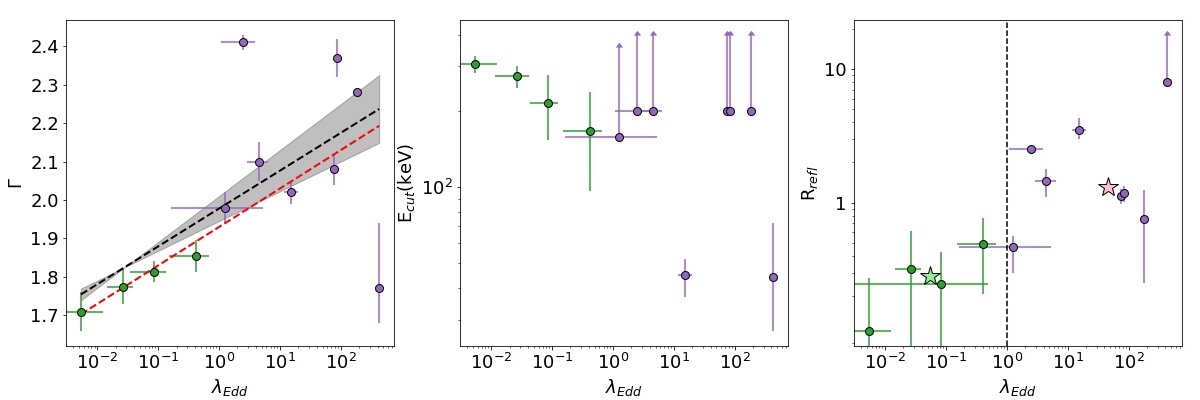}
    \caption{Top panels: Histogram of the density distributions of the photon-index ( $\Gamma$, left), of the high-energy cut-off ($E_c$ (center), and of the reflection fraction ($R_{\rm refl}$, right), for the SEAMBHs sources (purple) compared with the values of the Sy1 galaxies from the BASS sample (green) \citet{Ricci_2017,Ricci_2018}. Dashed lines represent the lower limits. Bottom panels: from left to right we plotted the $\Gamma-\lambda_{\rm Edd}$ relation, the $E_c-\lambda_{\rm Edd}$ relation and the $R_{\rm refl}-\lambda_{\rm Edd}$ relation of the SEAMBHs sources of our sample (purple) together with the median values of the Sy1 galaxies from the BASS sample (green)  \citep{Ricci_2017,Ricci_2018}. The dashed black line is the linear regressions while the shaded black regions represent the combined 3$\sigma$ error on the slope and normalisation. The dashed red line is the correlation found for 228 nearby ($0.01<z<0.5$) AGN belonging to the BASS sample from \citet{2017MNRAS.470..800T}. For the sources for which the $E_c$ is not constrained we used the value $=200$\,keV for consistency of what stated in Section \S \ref{sect:best_fit}. The two stars represent the median value for the BASS sample and for our sample of SEAMBHs.}
    \label{fig:primary_cont_param}
\end{figure*}
Using the best fitting values obtained from the best fitting models (see Section \S\ref{sect:best_fit}) we compared in the top panel of Figure \ref{fig:primary_cont_param} our analysis with the values obtained for the analysis of the Swift/BAT AGN Spectroscopical Survey (BASS\footnote{\url{www.bass-survey.com}}) BASS hard X-ray selected sample of unobscured ($N_H\leq10^{22}\rm cm^{-2}$) galaxies which includes sources with low and moderate accretion rate, $\lambda_{\rm Edd}\in [0.001-1.0]$ \citep{2017ApJ...850...74K,Ricci_2017,Ricci_2018}. We computed the median and the error of the photon-index, of the high-energy cut-off and of the reflection fraction of the BASS sources. To this end, we divided the Eddington ratio into 4 bins (as in \citealt{Ricci_2018}). These bins are not symmetrical, since we requested each bin to have at least 15 data values. We calculated the median and the error of the aforementioned spectral parameters in each Eddington ratio bin using the survival analysis method (SA). We used the \textsc{asurv} \citep{1985ApJ...293..192F} package, which applies the principles of SA to astronomical data. Information contained in the upper/lower limits can be then included in the measurement of statistical properties of samples without biasing results. Specifically, \textsc{asurv} calculates the non-parametric Kaplan-Meier product-limit (KMPL) estimator for a sample distribution. The KMPL estimator is an estimate of the survival function, which is simply 1-CDF (cumulative distribution function). Using the KMPL, we calculate for each Eddington ratio bin the median of the photon-index, of the high-energy cut-off and of the reflection fraction (50th percentile) and we estimate their uncertainty using the 16th and 84th percentiles. Then we used the SA median values of the BASS sources together with the values of the spectral parameters of the SEAMBHs of our sample to look for correlations with the Eddington ratio. We fitted the data applying the method of the "censored fit" \citep{Guainazzi2006,Bianchi2009} performing a large number of least square fits on a set of Monte-Carlo simulated data derived from the experimental points. In this method each detection was substituted by a random Gaussian distribution, whose mean is the best-fit measurement and whose standard deviation is its statistical uncertainty and each upper limit $U$ (lower limit $L$) was substituted by a random uniform distribution in the interval $[A,U]$ ($[L,A]$), where A is an arbitrary value $A \ll U$ ($A \gg L$).\\
We do not find any significant correlation between the high energy cut-off and the Eddington ratio and just a marginally statistically significant correlation between the reflection fraction and the Eddington ratio (see bottom central and left panels of Figure \ref{fig:primary_cont_param}). In fact we obtain a Pearson correlation coefficient of 0.09 corresponding to a $1-P_{\rm value}$ of 23\% for the $E_c-\lambda_{\rm Edd}$ relation and a Pearson correlation coefficient of 0.65 corresponding to a $1-P_{\rm value}$ of 96.8\% for the $R_{\rm refl}-\lambda_{\rm Edd}$ relation. Regarding the $\Gamma-\lambda_{\rm Edd}$ relation, we found, as expected, a robust correlation with a Pearson correlation coefficient of 0.83 corresponding to a $1-P_{\rm value}$ of 99.8\%. The presence of a strong and statistically significant positive correlation is well-known in literature \citep{1997AAS...190.5102B,2006ApJ...646L..29S,2008ApJ...682...81S,2009ApJ...700L...6R,2013MNRAS.433..648F,2013MNRAS.433.2485B,2016ApJ...826...93B,2016ApJS..225...14K,2017MNRAS.470..800T,Ricci_2018}. These past studies demonstrated that the main driver of the $\Gamma-\lambda_{\rm Edd}$ relation is the accretion rate \citep{2013MNRAS.433.2485B} and that with this relation is also possible to give an explanation to the X-ray Baldwin effect \citep{2013MNRAS.435.1840R}. We fitted the $\Gamma-\lambda_{\rm Edd}$ relation with the relation: $\Gamma=\alpha\log(\lambda_{\rm Edd})+\beta$ obtaining the following slope and intercept values: 
\begin{equation} 
\alpha=0.13 \pm 0.03 ;\, \beta=1.99 \pm 0.05
\label{eq:params}
\end{equation}
These values are consistent with what found by \citet{2013MNRAS.435.1840R} for a sample of 36 nearby AGN and for the BASS sample by \citet{2017MNRAS.470..800T,Ricci_2018} (see right bottom panel of Figure \ref{fig:primary_cont_param}). The correlation we found has different weight in the low and high accretion states, since we are considering single values for the sources with high Eddington ratios and median values resulting from SA for the sources with low/moderate Eddington ratios. Unfortunately the sample of sources accreting in super-Eddington accretion regime is not enough populated to carry on a survival analysis as it has been done for the sources accreting in the sub-Eddington regime. With this analysis we expanded this relation towards a wider range of accretion rate and we found that also sources accreting in the super-Eddington regime most likely follow the expected trend for the $\Gamma-\lambda_{\rm Edd}$ relation.\\
\begin{figure}
	\includegraphics[width=0.45\textwidth]{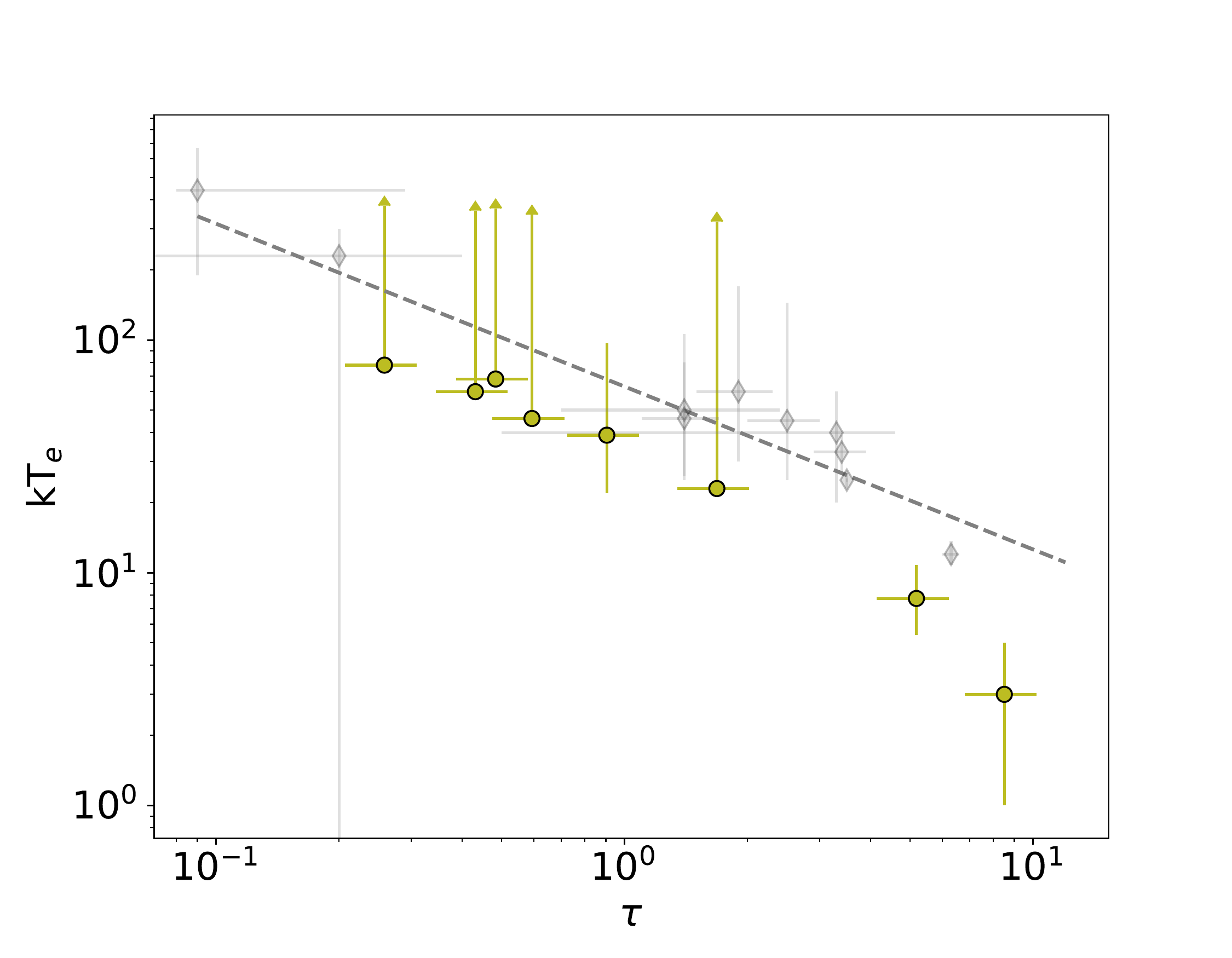}
    \caption{Plot of the coronal temperature (kT$_e$) versus the optical depth ($\tau$). The gray diamonds represent the values from \citet{2018A&A...614A..37T}}
    \label{fig:kt_t}
\end{figure}
We estimated the coronal temperature by fitting the spectra with the Models B3 and C3  (see Section \S \ref{sect:fitting_models}), and we estimated the optical depth using the relation from \citet{Pozdnyakov1997}:
\begin{equation}
\Gamma \sim 1+ \frac{\left[ \frac{2}{\Theta_e+3} -\log(\tau)\right]}{\log(12\Theta_e^2+25\Theta_e)}
\label{eq:gamma}
\end{equation}
where $\Gamma$ is the photon index of the spectrum between 2 and 10\,keV. The dependence from
 the optical depth is in the relativistic $y$ parameter is:
\begin{equation}
y=4(\Theta_e+4\Theta_e^2)\tau(\tau+1)
\label{eq:comptonparam}
\end{equation}
where $\Theta_e$ is the electron temperature normalized to the electron rest energy \citep{1994ApJ...429L..53G}:
\begin{equation}
\Theta_e=\frac{kT_e}{m_ec^2}
\label{eq:theta}
\end{equation}
Coronal temperature and optical depth values are reported in Table \ref{tab:coronal-param}. We use these values to compare the results obtained for super-Eddington sources with the values from \citet{2018A&A...614A..37T} finding that, even if most of the sources show a lower-limit on the coronal temperature, also super-Eddington sources appear to follow the same anti-correlation found for low-moderate accretion rates. The presence of an anti-correlation between the temperature of the AGN corona and its optical depth imply that either there are a geometrical variation in the disk-corona configuration (i.e., variation of the height of the corona above the accretion disk, $H$ or of the transition radius, $R_{t}$, separating the inner corona from the outer accretion disk) or in the intrinsic disk emission (i.e., the fraction of thermal emission due to viscous dissipation with respect to the total disk emission) \citep{2018A&A...614A..37T}. Objects with large corona temperatures would have a smaller $R_{t}$/$H$ or a larger contribution of the disk intrinsic emission than objects with lower temperature. %The disk emission arising from super-Eddington accretion flow is expected to be anisotropic and to be affected by photon trapping through electron scattering in the dense matter and advection cooling, thus the latter hypothesis is the most reliable to explain this anti-correlation in these sources. \\
To understand the various physical properties of a finite, thermal plasma it could be useful to look at the compactness-temperature diagram ($\Theta_e-\ell$) (\citealt{Svensson1984,Stern1995,Fabian2015,Fabian2017} and references therein). $\Theta_e$ is the aforementioned electron temperature and $\ell$ is the dimensionless compactness parameter:
\begin{equation}
\ell=\frac{L}{R}\frac{\sigma_T}{m_ec^3}
\label{eq:ell}
\end{equation}
where L is the luminosity, R is the radius of the corona and $\sigma_T$ is the Thompson cross section.\\
\begin{figure*}
	\includegraphics[width=0.7\textwidth]{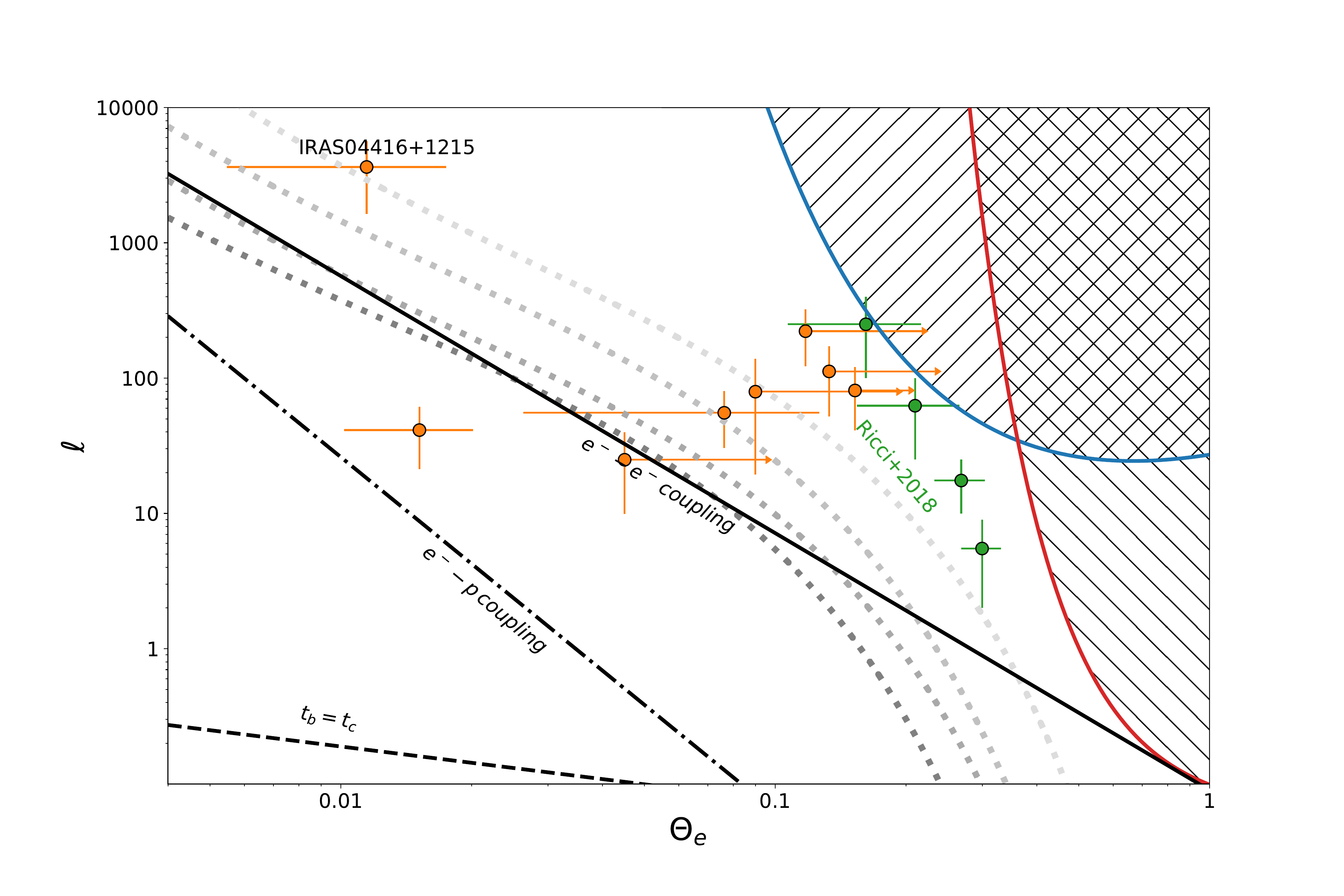}
    \caption{Theoretical compactness-temperature diagram. The black solid line represents the line below which the electron-electron coupling time scale is shorter than the Compton cooling time scale, the black dot-dashed line represents the line below which the electron-proton coupling time scale is shorter than the Compton cooling time scale while the black dashed line represents the line below which the dominant process is the Bremsstrahlung. The red and blue solid curves are the pair run-away lines respectively for a disk-like \citep{Stern1995} or a spherical corona \citep{Svensson1984}. The dotted gray curves are the $\Theta_e - \ell$ distribution for different values of the fraction of thermal, $\ell_{\rm th}$, and non-thermal, $\ell_{\rm nth}$, electrons, for a given ratio between the injection of soft photons ($\ell_{\rm s}$) and the total heating ($\ell_{\rm h}=\ell_{\rm th} + \ell_{\rm nth}$) of $\ell_{\rm s}/\ell_{\rm h}=0.1 $ [from darker to lighter: $\ell_{\rm nth}/\ell_{\rm th}=(0.33; 0.29; 0.23; 0.17)$, see \citealt{Fabian2017}]. The orange points represent the other sources of our sample, including IRAS\,04416+1215 \citep{dinosaur}. The green points represent the median values of the of compactness-temperature diagram for the BASS hard X-ray selected sample from \citet{Ricci_2018}.}
    \label{fig:compactness-temperature_diagram}
\end{figure*}
\begin{table}
\centering
\caption{Coronal parameters when fitting the X-ray broad-band \textit{XMM-Newton} plus \textit{NuSTAR} spectra of our sample of super-Eddington sources with a Comptonization model. Errors are at 90\% confidence levels.}
\begin{tabular}{lcccc}
\hline\hline
\noalign{\smallskip}
Source & Model$^{\star}$ &$kT_e [keV]$ & $\tau$& $\chi^2$/dof\\
\hline\hline
\noalign{\smallskip}
IRAS\,04416+1215 & B3 &$2.95 \pm 0.25$ &$8.50\pm1.70$& 1.08\\
\noalign{\smallskip}
IRASF\,12397+3333& C3 &$>60$&$0.45\pm0.09$ & 1.08 \\
\noalign{\smallskip}
Mrk\,493& C3 & $>68$&$0.48\pm0.09$ & 1.06\\
\noalign{\smallskip}
Mrk\,142& C3 & $>48$&$0.59\pm0.11$ & 1.08\\
\noalign{\smallskip}
Mrk\,382& C3 &$8 \pm 3$&$5.18\pm1.04$ &1.03\\
\noalign{\smallskip}
PG\,0026+129 & C3 & $39^{+58}_{-17}$&$0.90\pm0.18$ &1.09\\
\noalign{\smallskip}
PG\,0953+414& B3 & $>78$ &$0.25\pm0.05$& 1.09\\
\noalign{\smallskip}
NGC\,4748& B3 & $>23$ &$1.68\pm0.34$ & 0.87\\
\noalign{\smallskip}
\hline \hline
\end{tabular}\\
\label{tab:coronal-param}
\smallskip
{\raggedright \footnotesize{($\star$) The fitting models are described in Section \ref{sect:fitting_models} and listed in Table \ref{tab:models}.} \par}
\end{table}
The dominant radiation process in a plasma will be the one with the shortest cooling time. In the hot corona the most significant processes are Bremsstrahlung, the inverse Compton scattering and the pair production. Considering a spherical source
of size R and scattering optical depth $\tau$, which generates a luminosity L, the cooling time of the inverse Compton effect is :
\begin{equation}
t_C=\frac{3\pi R}{2c\ell(1+\tau)}
\end{equation}
while the Bremsstrahlung cooling time is 
\begin{equation}
t_{br}=\frac{\sqrt{\Theta}R}{\tau c \alpha_f}
\end{equation}
where $\alpha_f$ is the fine-structure constant. Comparing the two cooling times, it is possible to see that the Comptonization dominates at high compactness ($\ell>3\alpha_f(\Theta)^{-1/2}$). When $3\alpha_f\Theta<\ell<0.04\Theta^{-3/2}$ the dominant effect is the electron-proton coupling while for $0.04\Theta^{-3/2}<\ell<80\Theta^{-3/2}$ the electron electron coupling becomes relevant \citep{Fabian_1994}. Beyond a certain regime the pair production becomes a run-away process. In the $\Theta_e-\ell$ plane this regime is identified by the, so-called, pair runaway lines. The position of these lines depends on the shape of the source and on the radiation mechanism. \citet{Stern1995} computed the pair balance curve for a slab corona (red line in Figure \ref{fig:compactness-temperature_diagram}). \citet{Svensson1984} estimated that the pair balance for an isolated cloud occurs where $\ell\sim10\Theta^{5/2}e^{1/\Theta}$ (blue line in Figure \ref{fig:compactness-temperature_diagram}).\\
We compared the results obtained for our sample with the results of \citet{Fabian2015} and \citet{Ricci_2018} to see the location of  the super-Eddington AGN in the compactness-temperature diagram. We estimated the coronal temperature of the sources of our sample by fitting the spectra with the Model B3 and C3 (see Section \S \ref{sect:fitting_models} and Table \ref{tab:models}). To compute the compactness parameter of our sample we adopted the luminosity of the power-law component extrapolated to the 0.1--200\,keV band. Since from spectral-timing studies, such as reverberation and broad iron line fitting, yielded coronal sizes in the range of 3 to 10 gravitational radii ($R_g$) we assume these two limit values in the extrapolation of the compactness parameter \citep{2005MNRAS.359.1469M,2014ApJ...787...83M,Fabian2015}.\\
Even if most of the sources of our sample show an upper limit to the coronal temperature they are located close to the pair run-away line, suggesting that the AGN spectral shape is controlled by pair production and annihilation, as the majority of AGN \citep{Ricci_2018,Fabian2015,2014ApJ...787...83M}. The only two exceptions are IRAS\,04416+1215, which was extensively discussed in \citet{dinosaur}, and Mrk\,382 which shows features very similar to IRAS\,04416+1215. In fact this source show a low coronal temperature and a high optical depth. The location of Mrk\,382 in the $\theta-\ell$ diagram could suggests that its X-ray emission is generated in a thermal plasma which is not pair-dominated. However, its X-ray emission can be explained considering the scenario in which the corona is most likely made by an hybrid plasma, composed of both thermal and non-thermal electrons (e.g., \citealp{1993ApJ...414L..93Z,Fabian2017}). As in the case of IRAS\,04416+1215, a possible explanation can be found considering that, in super-Eddington accretions flows, the comptonizing corona could originate from radiation-pressure driven optically thick ($\tau \gtrsim 3$) outflows which act like a corona above the disc at relatively low temperatures ($kT_e \lesssim 10$\,keV, see \citealt{kawanaka2021determines}
\section{Conclusions}
\label{sect:conclusion}
Super-Eddington accretion is thought to play an important role in the co-evolution of SMBHs and their host galaxies (e.g.,\citealt{2013ARA&A..51..511K}), and to be the driving mechanism of the fast growth of the first SMBHs we observe in massive galaxies at high redshifts (e.g., \citealp{2012Sci...337..544V}). 
The study of super-Eddington accretion flows can also help our understanding of tidal-disruption events (e.g., \citealt{10.1093/mnras/sty971}) and of  Ultra-luminous X-ray sources (e.g., \citealt{begelman2006}). However, among all accretion modes, this phase is still the least understood one, and it is still largely debated what are the physical properties of the accretion flow and of the X-ray source for extreme accretion rates. Here we have presented the first systematic broad-band X-ray study of eight sources belonging to the SEAMBHs sample via simultaneous \textit{NuSTAR} and \textit{XMM-Newton} or \textit{Swift}/XRT observations, which are part of a dedicated campaign aimed at understanding the X-ray properties of these objects. All SEAMBHs objects have black hole masses estimated by reverberation mapping \citep{2015ApJ...806...22D}, which allows to accurately constrain the Eddington ratio.\\
The results of our work are the following:
\begin{list}{-}{\setlength{\itemsep}{0.cm}}
    \item The spectral shape of the sources of our sample is consistent with that of other NLS1s. The best-fitting model for most of the sources of our sample includes a soft-excess, ionized outflowing components, a primary continuum and a relativistic reflection component. 
    \item The super-Eddington accreting AGN of our sample show a primary continuum emission with a steep slope, $\Gamma\geq 2$, as expected for sources with high accretion rate \citep{2009Natur.459..540F,2014ApJ...795..147R}. The high energy cutoff is unconstrained for six sources, which show a lower limit, and it is well constrained for two sources: IRAS\,04416+1215 ($E_c=44^{+28}_{-17}$\,keV) and Mrk\,382 ($E_c=45^{+7}_{-8}$\,keV). Moreover, for most of the sources of our sample the reprocessed radiation is dominating over the primary continuum.
    \item For most of the sources of our sample the reprocessed radiation is dominating over the primary continuum. The median reflection fraction value of the SEAMBHs sources of our sample is a factor $\sim 2$ higher with respect to the median value of the reflection fraction of the Sy1 sources from the BASS sample.
    \item We found a correlation between the photon-index of the primary power-law and the Eddington ratio considering the values obtained for the SEAMBHs sources of our sample and the median values (extrapolated via survival analysis) of the unobscured AGN belonging to the BASS sample from \citet{Ricci_2018}. This correlation is consistent with past studies \citep{2013MNRAS.435.1840R,2017MNRAS.470..800T,Ricci_2018} and it corroborates the hypothesis that also the sources accreting in the super-Eddington regime show the expected trend for the $\Gamma-\lambda_{\rm Edd}$ relation.It should be noted that in calculating the correlation we are giving different weights to AGN accreting at low and high accretion states, since we are considering single values for the sources with high Eddington ratios and median values resulting from SA for the sources with low/moderate Eddington ratios.
    \item The Iron K$\alpha$ lines of the sources of our sample show relativistic broadening. Looking at the relation between the equivalent width of the narrow component of the FeK$\alpha$ with the 2--10\,keV flux, and with the Eddington ratio, the sources of our sample seem to show the ‘Iwasawa-Taniguchi’ effect, also known as 'X-ray Baldwin' effect, following the anti-correlation predicted by \citet{Shu2010} and \citet{Bianchi2007}, but only if we consider the equivalent width values extrapolated from a simple fitting model which includes a power-law and a narrow Gaussian line. From the fit with the best fitting model we found just upper limits to the Fe K$\alpha$ equivalent widths of the sources of this sample,suggesting that most of the Fe K$\alpha$ emission is in the broad, relativisitc component.
    \item We constrained the coronal temperature for three sources of our sample: Mrk\,382 (kT$_e=7.75^{+3.07}_{-2.35}$\,keV), PG\,0026+129 ($kT_e=39^{+58}_{-17}$\,keV) and IRAS\,04416+1215 (kT$_e=2.95\pm0.25$\,keV) while for the other sources we found just lower limits. Using these values together with the optical depth extrapolated using the relation from \citet{Pozdnyakov1997}, we found that the super-Eddington source of our sample seem to follow the anti-correlation between coronal temperature and optical depth found by \citet{2018A&A...614A..37T}. This implies that also in super-Eddington sources if the  disk-corona configuration is in radiative balance, there should be differences from source to source, in either the disk-corona configuration or in the intrinsic disk emission.%This is most likely due to the dramatically change in the accretion flow of the super-Eddington accreting sources with respect to standard accreting sources, due to photon trapping of electron scattering in the dense matter and advection cooling.
    \item Looking at the position of the SEAMBHs sources of our sample in the compactness-temperature diagram, most of the sources are located close to the pair run-away lines and above the e$^-$-e$^-$ coupling line, like most of the observed AGN \citep{Fabian2015,Ricci_2018}, suggesting that the AGN primary emission is controlled by pair production and annihilation. The location of IRAS\,04416+1215 and of Mrk\,382 in the $\theta-\ell$ diagram suggests that the X-ray emission of these sources is consistent with the scenario in which the corona is made of an hybrid plasma, i.e. composed both by thermal and non-thermal electrons. These sources show also a high value of optical depth. This is not surprising, considering that, in super-Eddington accretions flows, the comptonizing corona could originate from radiation-pressure driven optically thick ($\tau \gtrsim 3$) outflows which act like a corona above the disk at relatively low temperatures ($kT_e \lesssim 10$\,keV, see \citealt{kawanaka2021determines}).
\end{list}
The analysis performed in this work is applied over all the broad-band X-ray spectrum for each sources (0.3--25\,keV) including also all the spectral component present in the soft energy band, such as the soft excess and several warm absorbing components. The detailed analysis and spectral parameters values of the warm-absorbers, as well as the analysis of the light-curves of the sources of the sample, will be presented in a forthcoming paper (Tortosa et al. in prep.).

\section*{Acknowledgments}
AT acknowledge the support from FONDECYT Postdoctorado for the project n. 3190213. CR acknowledges support from the Fondecyt Iniciacion grant 11190831 and ANID BASAL project FB210003. LCH was supported by the National Science Foundation of China (11721303, 11991052, 12011540375) and the China Manned Space Project (CMS-CSST-2021-A04, CMS-CSST-2021-A06). This work is based on observations obtained with the ESA science mission \xmm, with instruments and contributions directly funded by ESA Member States and the USA (NASA), the \nustar mission, a project led by the California Institute of Technology, managed by the Jet Propulsion Laboratory and funded by NASA and the NASA \textit{Swift} mission. This research has made use of the \nustar Data Analysis Software (NuSTARDAS) jointly developed by the ASI Space Science Data Center (SSDC, Italy) and the California Institute of Technology (Caltech, USA). This research has made use of data obtained through the High Energy Astrophysics Science Archive Research Center Online Service, provided by the NASA/Goddard Space Flight Center. We acknowledge the use of public data from the \textit{Swift} data archive. The authors thank the referee for the useful suggestions which helped in improving the manuscript.\\ 

%%%%%%%%%%%%%%%%%%%%%%%%%%%%%%%%%%%%%%%%%%%%%%%%%%
\section*{Data Availability}
All the data utilized in this paper are publicly available in the \xmm data archive at \url{https://nxsa.esac.esa.int/nxsa-web/#search}, in the \nustar data archive at \url{https://heasarc.gsfc.nasa.gov/db-perl//W3Browse/w3browse.pl} and in the \textit{Swift} data archive at \url{https://swift.gsfc.nasa.gov/archive/}.\\ 
More details of the observations are listed in Table \ref{tab:observations_table}.


\bibliographystyle{mnras}
\bibliography{bibliography} 
%%%%%%%%%%%%%%%%%%%%%%%%%%%%%%%%%%%%%%%%%%%%%%%%%%

%%%%%%%%%%%%%%%%% APPENDICES %%%%%%%%%%%%%%%%%%%%%
\appendix
\section{Targets of the sample}
\label{app:targets}
The sources of our sample are part of the Super-Eddington Accreting Massive Black Holes sample (SEAMBHs, \citealt{Du2014,Wang2014a,2015ApJ...806...22D}) which includes objects with black hole masses estimated from reverberation mapping. We selected the sources with some of the highest Eddington ratios in the SEAMBHs sample (see Table \ref{tab:sources_table}). All the sources are Narrow Line Seyfert 1 galaxies accreting in a super-Eddington regime and their 2--10\,keV photon index is known to be relatively steep ($\Gamma_{2-10}>1.8$, e.g. \citealt{1999ApJS..125..317L,2004MNRAS.347..269G,2009Natur.459..540F,2014ApJ...795..147R,2014MNRAS.440.2347M}).\\
\indent \textbf{IRAS\,04416+1215} (z=0.0889) has the highest Eddington ratio in our sample and the highest known Eddington ratio in the local Universe. Its X-ray spectrum shows a multi-phase absorption structure, a prominent soft excess component and a hard X-ray emission dominated by the reprocessed radiation. Moreover, it shows the lowest coronal temperature measured so far by \nustar. A comprehensive investigation of the X-ray spectral properties of IRAS\,04416+1212 has been carried out in \citet{dinosaur}.\\
\indent \textbf{IRASF\,12397+3333} (z=0.0435) shows a steep X-ray spectrum ($\Gamma=2.05-2.30$; \citealp{Petrucci2018,Ursini2020}) as well as a soft excess in the soft X-ray energy band and  the presence of dusty ionized gas (Warm Absorbers) along the line of sight. \citep{1998A&A...333..827G,2007AAS...211.4418W};\\
\indent \textbf{Mrk\,493} is a local (z=0.0313) which shows, from previous \xmm observations, an X-ray spectrum with a prominent soft-excess and Iron K$\alpha$ emission line \citep{1985ApJ...297..166O,2019ApJ...870L..13A}. Its X-ray spectrum was well fitted by a blurred reflection model that is reflection dominated and the blurred reflector is also needed to fit the soft excess component \citep{2018MNRAS.477.3247B};\\
\indent \textbf{Mrk\,142} (z=0.0449) is a well-studied reverberation-mapped AGN \citep{2020ApJ...896....1C}. From the study of the Lijiang telescope monitoring in 2012-2013, \citet{Li_2018} found that the Broad Line region of Mrk\,142 is described by a disk with inclination angle of $42^{\circ}$.\\
\indent \textbf{Mrk\,382} (z=0.0337) is a bright local AGN detected in the X-ray for the first time by the European X-Ray Observatory SATellite (EXOSAT) in 1983. From the analysis of this observation, Mrk\,382 comes out to have a steep primary continuum with $\Gamma=1.72\pm0.14$ \citep{1992A&A...262...49S}; \\
\indent \textbf{PG\,0026+129} (z=0.142) is a bright radio-quiet quasar. It has been observed in the X-ray band by EXOSAT in 1983 and 1984. The EXOSAT spectrum was described by an absorbed power-law with spectral index 1.5 \citep{1988ApJ...330..178T}. \\
\indent \textbf{PG\,0953+414} (z=0.234) is a low-redshift radio-quiet quasar observed for the first time by the Advanced Satellite for Cosmology and Astrophysics (ASCA). It showed a spectrum well fitted by an absorbed power-law model with a photon index $2.03\pm0.05$ \citep{2000ApJ...531...52G}. Analysing a \xmm observations, \citet{2004A&A...422...85P} found that the broad-band continuum could be well fitted either by a broken power-law model or by a double Comptonization model. The source showed  the presence of a soft excess and the 2-5\,keV spectrum was well fitted by an absorbed power-law model with $\Gamma=2.04 \pm 0.11$. \\
\indent \textbf{NGC\,4748} (z=0.0146) is also known as MCG\,-02-33-034. It has a steep spectrum with a photon-index $\Gamma=2.50 \pm 0.17$ \citep{2001A&A...368..797P} and $\Gamma=2.20\pm0.11$ according to \textit{Swift}/XRT data \citep{2010MNRAS.403..945L}. \citet{2011MNRAS.417.2426P} from the analysis of \textit{INTEGRAL/IBIS} data found for this source a spectrum with $\Gamma=2.01\pm0.13$ and no high energy cut-off, no Iron K$\alpha$ emission line and no reflection component. Using the \xmm (EPIC and OM), \textit{Swift}/BAT and INTEGRAL/ISGRI observations \citet{Vasylenko2018} found the presence of a strong ionized reflection that could be responsible for the soft-excess component.
\begin{table}
\caption{Summary of the simultaneous observations of the sources of our sample. The Net count rate is extrapolated in the 0.3--10\,keV range for EPIC cameras and in the 3--25\,keV interval for both FPMA and B.}
\label{tab:observations_table}
\begin{tabular}{lccc}
\hline
\hline
 & \xmm & \multicolumn{2}{c}{\nustar}\\
 & Epic-pn & FPMA& FPMB\\
\hline
\hline
&\multicolumn{3}{c}{\textbf{IRAS\,04416+1215}}\\
\hline
OBS.ID & 0852060101 & \multicolumn{2}{c}{60560026002} \\
Exposure Time (ks) & 81.1  & 76.9 & 76.9 \\
Net Exposure Time (ks) & 56.5 & 71.3 & 71.3 \\
Net Count Rate ($\rm{counts}/s$)& 0.477 & 0.026 & 0.025\\
\hline
&\multicolumn{3}{c}{\textbf{IRASF\,12397+3333}}\\
\hline
OBS.ID & 0843020601 & \multicolumn{2}{c}{60501007002} \\
Exposure Time (ks) & 58.0  & 48.7 &48.7 \\
Net Exposure Time (ks) & 36.2 & 48.3 & 48.3\\
Net Count Rate ($\rm{counts}/s$)& 7.60 & 0.14 & 0.13\\
\hline
&\multicolumn{3}{c}{\textbf{Mrk\,493}}\\
\hline
OBS.ID &0852060201 & \multicolumn{2}{c}{60560027002} \\
Exposure Time (ks) &69.7   &62.8  &62.8\\
Net Exposure Time (ks) & 44.4&62.5&62.4\\
Net Count Rate ($\rm{counts}/s$)&2.92&0.043&0.040\\
\hline
&\multicolumn{3}{c}{\textbf{Mrk\,142}}\\
\hline
OBS.ID &0852060301& \multicolumn{2}{c}{60560028002} \\
Exposure Time (ks) & 64.0 &75.2  &75.2 \\
Net Exposure Time (ks) &43.5&74.2&74.5 \\
Net Count Rate ($\rm{counts}/s$)&3.98&0.027&0.028 \\
\hline
&\multicolumn{3}{c}{\textbf{Mrk\,382}}\\
\hline
OBS.ID &0843020801 & \multicolumn{2}{c}{60501008002} \\
Exposure Time (ks) & 52.9 &52.3  &52.3 \\
Net Exposure Time (ks) &35.7&52.2&52.0\\
Net Count Rate ($\rm{counts}/s$)&0.63&0.027&0.032\\
\hline
\hline
\bigskip\\
\hline
\hline
& \textit{Swift} & \multicolumn{2}{c}{\nustar}\\
 & XRT& FPMA& FPMB\\
\hline
\hline
&\multicolumn{3}{c}{\textbf{PG\,0026+129}}\\
\hline
OBS.ID &00089100001& \multicolumn{2}{c}{60663003002} \\
Exposure Time (ks) &  9.8&147.3  &147.3 \\
Net Exposure Time (ks) &9.7&147.2&146.0 \\
Net Count Rate ($\rm{counts}/s$)&0.32&0.11&0.11 \\
\hline
&\multicolumn{3}{c}{\textbf{PG\,0953+414}}\\
\hline
OBS.ID &00089098003 & \multicolumn{2}{c}{0663001002} \\
Exposure Time (ks) &4.7 &67.8  &67.8 \\
Net Exposure Time (ks) &4.6 &65.1&64.7\\
Net Count Rate ($\rm{counts}/s$)&0.27&0.089&0.082 \\
\hline
&\multicolumn{3}{c}{\textbf{NGC\,4748}}\\
\hline
OBS.ID & 00089099001& \multicolumn{2}{c}{60663002002} \\
Exposure Time (ks) & 4.7 &80.6  &80.6 \\
Net Exposure Time (ks) &4.6&67.2&68.1 \\
Net Count Rate ($\rm{counts}/s$)&0.38&0.11 &0.11 \\
\hline\hline
\end{tabular}
\end{table}
\section{Detailed spectral analysis of the sources of the sample}
\label{app:sources_details}
We performed a comprehensive spectral analysis of the sources of our sample using all the fitting models described in Section \ref{sect:fitting_models}. We report the details below:
\subsection*{IRAS 04416+1215}
For the details about the analysis of this source we refer to \citet{dinosaur} in which the detailed analysis of this source is carefully described.
\subsection*{IRASF 12397+3333}
In the soft X-ray band, IRASF\,12397+3333 shows the presence of a black body component for the soft excess with temperature kT$_{\rm bb}=90 \pm 8$\,eV and one ionized absorption components with turbulent velocity of 100\,$\rm km\,s^{-1}$. \\
We started the analysis using the phenomenological Model A. This model yielded a very good fit with $\chi^2$ = 1616 for 1509 degrees of freedom (dof). The source show a steep primary continuum with the photon index of the primary power-law being $\Gamma=2.44 \pm 0.03$ with an unconstrained the high energy cut-off $E_{\rm cut} >119$\,keV. We found a reflection fraction of $R_{\rm refl}=0.54 \pm 0.08$. The Iron K$\alpha$ emission line appears to be weak, showing a flux of $2.87 \pm 0.55 \times 10^{-6}\,\rm ph\, \rm cm^{-2}\, \rm s^{-1}$.\\ Switching to Model B1 we obtained an excellent fit $\chi^2$/dof = 1621/1511. The photon index was almost similar to the one obtained with Model A: $\Gamma=2.28\pm0.02$ but the reflection fraction value and was slightly lower, $R_{\rm refl}=0.30 \pm 0.11$. The ionization parameter was \logxi$=3.36\pm0.01$ and we found an Iron abundance value A$_{\rm Fe}=1.19\pm0.37$. With this model the cut-off energy was unconstrained, resulting in a lower limit of 740\,keV. The fit with Model C1 gave a better $\chi^2$ value: $\chi^2$/dof=1592/1510. With this model we obtained a constraint for the spin parameter $\mathit{a}=0.78^{+0.18}_{-0.29}$, the Iron abundance was lower with respect to Model B1: A$_{\rm Fe}=0.93^{+0.21}_{-0.14}$. The cut-off energy is still unconstrained, with a lower limit of $E_c>790$\,keV and the reflection fraction is $R_{\rm refl}=0.75^{+0.52}_{-0.38}$. Including the narrow component of the Iron line (Model C1+NL) it shows a flux of $1.12 \pm 0.45 \times 10^{-6}\,\rm ph\, \rm cm^{-2}\, \rm s^{-1}$, consistent with the values we found with Model A.\\
Then, we tested the different flavours of \textsc{xillver} and \textsc{relxill} models by fitting the spectra with the Models B2, B3, C2 and C3. The spectral parameters values obtained with Models B2 and C2 were exactly the same obtained with Models B1 and C1. The disk column density was unconstrained in both cases, and we obtained $\log(n / \rm cm^{-3})$<15.50 and $\log(n / \rm cm^{-3})$<15.58 respectively. The goodness of the fit with these Models worsen, the fits yielded a $\chi^2$/dof=1631/1511 (Model B2) and $\chi^2$/dof=1597/1510 (Model C2). Testing the reflection models given for an incident spectra with \textsc{nthcomp} Comptonization model (Model B3 and C3) we found a lower limit for the coronal temperature of $kT_e >120$\,keV (B3) and $kT_e >60$\,keV (C3). Also in these cases the fits are worst with respect to Models B1 and C1 with $\chi^2$/dof=1632/1511 for Model B3 and $\chi^2$/dof=1677/1510 for Model C3.\\
When applying the Model C4 ($\chi^2$/dof=1608/1511) to the broad-band spectra we found an inner index emissivity value of q$_1=8.84\pm0.33$ and a break radius value of $R_{\rm br}=5.63\pm0.22\, \rm R_g$. The value of the inner emissivity index seems extreme but it is needed to reproduce steepness of the soft excess spectrum. The break radius can be considered as a lower limit for the size of the corona. Whit this model the spin value is $\mathit{a}=0.89\pm0.11$ and an Iron abundance, A$_{\rm Fe}=0.60\pm 0.15$.\\
Lastly we fitted the spectra with Model D. The addition of the unblurred component to Model C1 provided a fit with $\chi^2$/dof=1593/1506. The high energy cut-off is still unconstrained, $E_c<180$\,keV and the photon index value is consistent with the values obtained with the other models: $\Gamma=2.29\pm0.04$. The reflection fraction of the the relativistic disk reflection component is $R_{\rm rel,refl}=0.27 \pm 0.16$ while the reflection fraction of the distant neutral reflection component is unconstrained $R_{\rm refl}<0.04$. Even if the statistical goodness of this fit is the best of all the tested models, the fact that it gives only an upper limit on the reflection fraction of the distant neutral component, it suggests that this component is much likely not present.
\subsection*{Mrk 493}
The soft excess component of Mrk\,493 shows a temperature of kT$_{\rm bb}=75 \pm 4$\,eV. In the soft energy band it shows also the presence of one ionized absorption component, one with turbulent velocity 100\,$\rm km\,s^{-1}$.\\
The first fit with the phenomenological Model A ($\chi^2$/dof=1002/970) resulted in a steep continuum, $\Gamma =2.42 \pm 0.03$ with a high energy cut-off $E_c= 104 \pm 32$\,keV which appeared to be reflection dominated: $R_{\rm refl}=1.32 \pm 0.08$. The Iron K$\alpha$ emission line flux was $9.96 \pm 0.58 \times 10^{-7}\,\rm ph\, \rm cm^{-2}\, \rm s^{-1}$.\\
Then next step was fitting the Mrk\,493 spectra with the Model B1. We obtained a very good fit: $\chi^2$/dof=1080/972. With this model the photon index of the primary continuum is flatter with respect to the one obtained with Model A, but the spectrum is still steep as expected for this kind of sources: $\Gamma=2.11\pm0.02$. Also the reflection fraction value and was lower, $R_{\rm refl}=0.41 \pm 0.07$. We found an Iron overabundance: A$_{\rm Fe}=4.78\pm1.02$ and ionization parameter value of \logxi$=2.96\pm0.06$. The high energy cut-off was not constrained with this model, we found a lower limit of 519\,keV. Switching to Model C1 we obtained a better fit: $\chi^2$ value: $\chi^2$/dof=1030/979. The spin parameter was unconstrained but we obtained an upper limit of $\mathit{a}<0.8$ for a counterclockwise rotation. The Iron abundance was still high: A$_{\rm Fe}=3.86^{+1.04}_{-1.01}$ and the other spectral parameters were almost the same and the cut-off energy show a lower limit of $E_c>827$\,keV. Including the narrow component of the Iron line (Model C1+NL) it shows a flux of $1.16 \pm 0.60 \times 10^{-6}\,\rm ph\, \rm cm^{-2}\, \rm s^{-1}$. The flux was consistent with the values we found with Model A.\\
Testing the different flavours of \textsc{xillver} and \textsc{relxill} models by fitting the spectra with the Models B2, B3, C2 and C3 we found the spectral parameters values obtained with Models B2 and C2 were almost similar to the values obtained with Models B1 and C1. The disk column density was unconstrained in both cases, we found $\log(n / \rm cm^{-3})$<16.68 and $\log(n / \rm cm^{-3})$<15.51 respectively. The goodness of the fit with these models worsen, the fits yielded a $\chi^2$/dof=1085/972 (Model B2) and $\chi^2$/dof=1033/971 (Model C2). Testing the reflection models given for an incident spectra with \textsc{nthcomp} Comptonization model (Model B3 and C3) we found a lower limit for the coronal temperature of $kT_e >81$\,keV (B3) and $kT_e >68$\,keV (C3). Also in these cases the fits are worst with respect to Models B1 and C1 with $\chi^2$/dof=1089/972 for Model B3 and $\chi^2$/dof=1034/971 for Model C3.\\
When fitting the spectra of Mrk\,493 with Model C4 ($\chi^2$/dof=1030/971), which accounts for the soft X-ray excess via a broken power-law emissivity function, we found a break radius value of $R_{\rm br}=4.85_{-0.86}^{+0.18}$ and the inner emissivity index q$_1=7.44_{-0.83}^{+0.86}$ and we are also able to constrain the spin value, which resulted to be $\mathit{a}=0.94\pm0.01$.\\
We tested the presence of a distant reflection component with Model D. The photon index value is consistent with the values obtained with the other models: $\Gamma=2.30\pm0.02$ and the high energy cut-off is still unconstrained, $E_c<866$\,keV. The reflection fraction of the distant neutral reflection component is unconstrained $R_{\rm refl}<1.92$. The statistical goodness of this fit is as good as for the fit with Model C1: $\chi^2$/dof=1065/968, but since the reflection fraction of the distant neutral component is unconstrained we do not consider it as a best-fitting model.
\subsection*{Mrk 142}
The source shows in the soft energy band a soft excess component with a temperature of kT$_{\rm bb}=90 \pm 9$\,eV and one ionized absorption component with turbulent velocity of 1000\,$\rm km\,s^{-1}$.\\
The fit with Model A yielded a good fit $\chi^2$/dof=408/357. With this model the broad-band spectra of the source is characterized by a steep power-law, $\Gamma=2.79\pm0.05$ with a high energy cut-off of $E_c=28^{+36}_{-18}$\,keV but we found just a lower limit for the reflection fraction of the reprocessed emission: $R_{\rm refl}<0.92$. The Iron K$\alpha$ line had a flux of $8.49 \pm 0.49 \times 10^{-7}\,\rm ph\, \rm cm^{-2}\, \rm s^{-1}$.\\
The next test was with Model B1 which gave a fit with $\chi^2$/dof=398/359. For the primary power-law we found $\Gamma=2.35\pm0.02$ with an unconstrained high energy cut-off: $E_c>385$\,keV. The reflection component had a reflection fraction value of $R_{\rm refl}=0.52 \pm 0.09$ and we found also an Iron overabundance (A$_{\rm Fe}>6.70$). Testing the relativistic reflection with model C1 ($\chi^2$/dof=386/358) we found almost the same parameters as before, the power-law photon index was $\Gamma=2.37\pm0.05$ and the high energy cut-off was still unconstrained with a lower limit of $E_c>763$\,keV. We found a constraint for spin parameter: $\mathit{a}=0.1\pm0.59$ for a counterclockwise rotation. Including the
narrow component of the Iron line (Model C1+NL) it shows a flux of  $1.47 \pm 0.54 \times 10^{-6}\,\rm ph\, \rm cm^{-2}\, \rm s^{-1}$, consistent with the values we found with Model A.\\ 
The fit with Models B2 and C2 were slightly worse with respect to the ones with Models B1 and C1: $\chi^2$/dof=399/359 and $\chi^2$/dof=389/358 respectively. We found only an upper limit for the disk column density, $\log(n / \rm cm^{-3})$<16.28 (Model B2) and $\log(n / \rm cm^{-3})$<16.49 (Model C2), while the other spectral parameters were almost the same as Models B1 and C1. The statistical goodness of the fit worsen when fitting the data with Model B3 and C3 ($\chi^2$/dof=403/359 and $\chi^2$/dof=388/358 respectively) and we found a lower limit to the coronal temperature: $kT_e >48$\,keV for both B3 and C3 Models. Applying Model C4 to the data ($\chi^2$/dof=376/358) we found a lower limit for the inner index emissivity value of q$_1>8.551$ and a break radius value of $R_{\rm br}=3.25^{+0.48}_{-0.38}$. Whit this model we constrained a spin value which resulted to be $\mathit{a}=0.987\pm0.002$. The last tested model was Model D which included two reflection components, one blurred (\textsc{relxill} component) and one neutral and unblurred (\textsc{xillver} component). The fit was very good,  $\chi^2$/dof=357/354, but, as for the previous sources, also in this case the reflection fraction of the neutral distant component is unconstrained, showing only an upper limit of $R_{\rm refl}<0.09$.
\subsection*{Mrk 382}
In the soft energy band the spectra of this source shows two warm absorbing components, one with turbulent velocity of 100\,$\rm km\,s^{-1}$ and one with turbulent velocity of 1000\,$\rm km\,s^{-1}$ plus a soft excess component with a temperature of kT$_{\rm bb}=25 \pm 5$\,eV.\\
The statistical goodness of Model A was very good $\chi^2$/dof=304/309. Fitting with this model we obtained for the photon index of the primary power-law a value $\Gamma=2.19\pm0.09$. The high energy cut-off is unconstrained with this model: $E_c> 43$\,keV as well as the reflection fraction: $R_{\rm refl}<0.77$. We found a flux of $2.09 \pm 0.57 \times 10^{-6}\,\rm ph\, \rm cm^{-2}\, \rm s^{-1}$ for the Iron K$\alpha$ line.\\ 
After testing the data with the phenomenological Model A, we tried out the different flavours of \textsc{xillver} and \textsc{relxill}. The first Model tested was B1 ($\chi^2$/dof=309/311). We obtained with this test a photon index parameter value of $1.93 \pm 0.07$, a cut-off energy of $62^{+45}_{-16}$\,keV and a reflection fraction $R_{\rm refl}=0.81 \pm 0.21$. Moreover we found an Iron abundance of A$_{\rm Fe}=2.02 \pm 0.93$. With the Model C1 we obtained a good fit, with $\chi^2$/dof=319/315. The photon index of the primary power-law was $\Gamma=2.02\pm0.03$ and we were able to measure the high energy cut-off: $E_c=45^{+7}_{-8}$ keV. We found a higher value of the reflection fraction compared to Model B1, $R_{\rm refl}=3.52^{+0.54}_{-0.76}$ but compatible within the errors. The spin parameter was unconstrained but we obtained a lower limit of $\mathit{a}>0.89$. Adding the narrow line component for the Iron K$\alpha$ line to Model C1 (Model C1+NL) we found a flux of $1.59 \pm 0.68 \times 10^{-6}\,\rm ph\, \rm cm^{-2}\, \rm s^{-1}$. The fits with Models B2 and C2 had the same statistical significance of the fits with Models B1 and C1 but we could not constrain the disk column density. We found $\log(n / \rm cm^{-3})$<15.61 and $\log(n / \rm cm^{-3})$<15.81 respectively. 
When fitting Mrk\,382 spectra with the reflection models given for an incident spectra with \textsc{nthcomp} Comptonization model (Model B3 and C3) we found a lower limit for the coronal temperature with Model B3 of $kT_e >12$\,keV and a measure with Model C3: $kT_e=7.75_{-2.35}^{+3.07}$\,keV. The fits in these tests are really good, with $\chi^2$/dof=304/311 for Model B3 and $\chi^2$/dof=325/315 for Model C3. When fitting the spectra of Mrk\,382 with Model C4 ($\chi^2$/dof=299/312), which accounts for the soft X-ray excess via a broken power-law emissivity function, we found a good constraint for the spin value: $\mathit{a}=0.91\pm0.06$, the break radius value was $R_{\rm br}=4.80_{-1.39}^{+2.69}$ and the inner emissivity index was q$_1=6.42\pm0.88$. With Model C4 we found a cut-off value of $E_c=38^{+15}_{-10}$ keV and a photon index $\Gamma=2.09\pm0.05$, however, we were not able to constrain the reflection fraction which shows just lower limit $R_{\rm refl}>8.16$. We also tested the presence of an unblurred neutral reflection component with Model D. This model provided a fit with $\chi^2$/dof=293/305. The reflection fraction values of the blurred and unblurred reflection components were $R_{\rm refl}=2.50\pm 0.96$ and $R_{\rm refl}=2.69 \pm 1.65$, respectively but the flux of the unblurred reflection component shows only an upper limits: F$_{\rm xill}<2.26 \times 10^{-6}\,\rm ph\, \rm cm^{-2}\, \rm s^{-1}$.
\subsection*{PG 0026+129}
%In the soft X-ray band of the spectra we found that PG\,0026+12 show a black body component with a temperature of kT$_{\rm bb}=155 \pm 7$\,eV and two ionized absorption components, one with turbulent velocity of 100\,$\rm km\,s^{-1}$ and one with turbulent velocity 1000\,$\rm km\,s^{-1}$. 
We started the analysis using the phenomenological Model A. This model yielded a very good fit with $\chi^2$ = 810 for 761 dof. The primary X-ray component is characterized by a photon index value of $\Gamma=1.93\pm 0.07$ and a cutoff energy of $E_c=61^{+50}_{-18}$. We found a lower limit for the reflection fraction of the neutral reflection component: $R_{\rm refl}>1.96$. Since \nustar does not have a sufficient effective area below 10\,keV and \textit{Swift}/XRT spectrum is background dominated above 3\,keV, it was not possible to detect the Iron K$\alpha$ emission line.\\
The fit with Model B1 gave a $\chi^2$ = 797 for 763 dof. The photon index of the primary power-law was $\Gamma=2.04 \pm 0.05$ and we found a cutoff energy of $E_c=137^{+150}_{-60}$\,keV. The reflection fraction with this model was constrained and it was $R_{\rm refl}=1.02 \pm 0.35$. Moreover we found an ionization fraction parameters of \logxi$=1.46^{+1.41}_{-0.43}$. With Model C1 the fit is very good ($\chi^2$/dof=778/762). We found a similar value for the photon index of the primary power-law as in Model B1 and moreover a lower limits on the spin of the black hole $\mathit{a}>0.65$ for a counterclockwise rotation. The high energy cut-off show a lower limit of $E_c>626$\,keV and the reflection fraction value was $R_{\rm refl}=1.45\pm0.35$. With this model we found an ionization fraction parameters of \logxi$=2.50\pm0.19$ However, when fitting with this model, PG\,0026+129 show an upper limit for the abundance of iron, in contrast to what found for the other sources which presented an iron overabundance, A$_{\rm Fe}<0.54$. We tried the fit with Model C1+NL, adding the narrow line component for the Iron K$\alpha$ line to Model C1, but the equivalent width for this component is consistent with zero most likely due to the fact that the \textit{Swift}/XRT spectrum is background dominated over 3\,keV plus the fact that \nustar sensitivity drops at energies below 10\,keV. When we applied Model B2 and C2 ($\chi^2$ was 782 for 763 dof and 778 for 762 dof respectively) we found a value of the disk density of $\log(n / \rm cm^{-3})=16.06^{+0.34}_{-0.29}$ in Model B2 and just an upper limit of $\log(n / \rm cm^{-3})<16.99$ in Model C2. The other spectral parameters are the same within Models B1 and B2 and Models C1 and C2 except for the photon index of the primary power-law of Model B2 which was steeper with respect to the value of Model B1: $\Gamma =2.21 \pm 0.06$. The statistical goodness of the fit worsen when fitting the data with Model B3 and C3 ($\chi^2$/dof=795/763 and $\chi^2$/dof=394/355 respectively) and we found a constraint for the coronal temperature: $kT_e=34_{-14}^{+102}$\,keV when applying Model B3 and $kT_e=39_{-17}^{+58}$\,keV when applying Model C3. Applying Model C4 to the data ($\chi^2$/dof=834/762) we found a lower limit for the inner index emissivity value of q$_1>8.76$ and a break radius value of $R_{\rm br}=3.75 \pm 0.59$. Whit this model we obtained a very good constraint for the spin value which resulted to be $\mathit{a} = 0.88^{+0.09}_{-0.03}$. The last tested model was Model D which included two reflection components, one blurred (\textsc{relxill} component) and one neutral and unblurred (\textsc{xillver} component). The fit was very good, $\chi^2$/dof=826/758, but, as for the previous sources, also in this case the reflection fraction of the neutral distant component is unconstrained, showing only an upper limit of $R_{\rm refl}< 0.002$.
\subsection*{PG 0953+414}
%In the soft energy band PG\,0953+414 does not show a prominent soft excess component but it show some features due to an ionized absorption component with turbulent velocity 1000\,$\rm km\,s^{-1}$. 
The first fit with the phenomenological Model A ($\chi^2$/dof=596/494) resulted in a steep continuum, $\Gamma =2.55 \pm 0.03$ with a high energy cut-off $E_c= 85_{-29}^{+118}$\,keV. We found an upper limit for the reflection component: $R_{\rm refl}<2.67 \pm 0.08$. Since \nustar does not have a sufficient effective area below 10\,keV and \textit{Swift}/XRT spectrum is background dominated above 3\,keV, it was not possible to detect the Iron K$\alpha$ emission line.\\
Then next step was fitting the PG\,0953+414 spectra with the Model B1. We obtained a very good fit: $\chi^2$/dof=531/494. With this model the photon index of the primary continuum was: $\Gamma=2.41\pm0.02$ and the reflection fraction value and was $R_{\rm refl}=2.44 \pm 0.18$. In contrast to what found for the other sources, we found an upper limit on the Iron abundance: A$_{\rm Fe}<0.53$. The ionization parameter value was \logxi$=1.45\pm0.35$. The high energy cut-off was not constrained with this model, we found a lower limit of 830\,keV. Switching to Model C1 we obtained a slightly worst fit: $\chi^2$ value: $\chi^2$/dof=538/493 and the spin parameter was consistent with zero.\\
Testing the different flavours of \textsc{xillver} and \textsc{relxill} models by fitting the spectra with the Models B2, B3, C2 and C3 we found the spectral parameters values obtained with Models B2 and C2 were almost similar to the values obtained with Models B1 and C1. We found a good constraint for the disk column density with Model B2:  $\log(n / \rm cm^{-3})=16.28_{-0.14}^{+0.47}$ and an upper limit with Model C2: $\log(n / \rm cm^{-3})$<16.58. The goodness of the fit with these models improved, the fits yielded a $\chi^2$/dof=528/494 (Model B2) and $\chi^2$/dof=528/493 (Model C2). Testing the reflection models given for an incident spectra with \textsc{nthcomp} Comptonization model (Model B3 and C3) we found a lower limit for the coronal temperature of $kT_e >78$\,keV (B3) and $kT_e >186$\,keV (C3). In these cases the fits are worst with respect to Models B1 and C1 with $\chi^2$/dof=540/494 for Model B3 and $\chi^2$/dof=459/493 for Model C3.\\
Since the spectra of PG\,0953+414 do not show a prominent soft excess component we do not tested the Model C4 on this source. We did not test Model D either since when fitting with Model C1 we excluded the presence of relativistic reflection.
\subsection*{NGC 4748}
%In the soft energy band NGC\,4748 show a soft excess component with a temperature of kT$_{\rm bb}=108 \pm 30$\,eV and two ionized absorption components, one with turbulent velocity of 100\,$\rm km\,s^{-1}$ and one with turbulent velocity 1000\,$\rm km\,s^{-1}$. 
We tested st first the phenomenological Model A. This first fit ($\chi^2$/dof=531/494) resulted in a steep continuum, $\Gamma =1.93 \pm 0.05$ with a high energy cut-off which show a lower limit of  $E_c>152$\,keV and the reflection fraction which show an upper limit of: $R_{\rm refl}<0.12$. Moreover the Iron K$\alpha$ line shows a flux of $4.29 \pm 0.97 \times 10^{-6}\,\rm ph\, \rm cm^{-2}\, \rm s^{-1}$.\\
Then we fitted NGC\,4748's spectra with the Model B1. We obtained a very good fit: $\chi^2$/dof=418/476 is . With this model the photon index of the primary continuum was: $\Gamma=1.98\pm0.04$ and the high energy cut-off show a lower limit of $E_c>158$\,keV while the reflection fraction value and was $R_{\rm refl}=0.47^{+0.17}_{-0.09}$. We found the iron abundance being A$_{\rm Fe}=1.87^{+1.12}_{-1.27}$. The ionization parameter value was \logxi$=1.20\pm0.96$. Switching to Model C1 we obtained a slightly worst fit: $\chi^2$ value: $\chi^2$/dof=419/475 and the spin parameter was consistent with zero.\\
Then we tested the different flavours of \textsc{xillver} and \textsc{relxill} models by fitting the spectra with the Models B2, B3, C2 and C3. When applying Model B2 and C2 to the NGC\,4748 spectra the goodness of the fit improved, the fits yielded a $\chi^2$/dof=417/476 (Model B2) and $\chi^2$/dof=418/475 (Model C2) and we found the spectral parameters values obtained were similar to the values obtained with Models B1 and C1. We foundan upper limit for the disk column density with both models: $\log(n / \rm cm^{-3})<17.22$ (Model B2) and  $\log(n / \rm cm^{-3})<17.87$ (Model C2). Testing the reflection models given for an incident spectra with \textsc{nthcomp} Comptonization model (Model B3 and C3) we found a lower limit for the coronal temperature of $kT_e>23$\,keV (B3) and $kT_e >16$\,keV (C3). In these cases the statistical goodness of the fits is the same as with Models B1 and C1 with $\chi^2$/dof=418/476 for Model B3 and $\chi^2$/dof 419/475 for Model C3.\\
Applying Model C4 to the data ($\chi^2$/dof=422/475) we found a lower limit for the inner index emissivity value of q$_1>8.16$ and a break radius value of $R_{\rm br}=4.56^{+0.78}_{-0.85}$ but again also with this model the spin parameter is consistent with being zero, thus, given the aforementioned results, we did not test Model D since we can excluded the presence of relativistic reflection.
\begin{figure*}
	\includegraphics[scale=0.372]{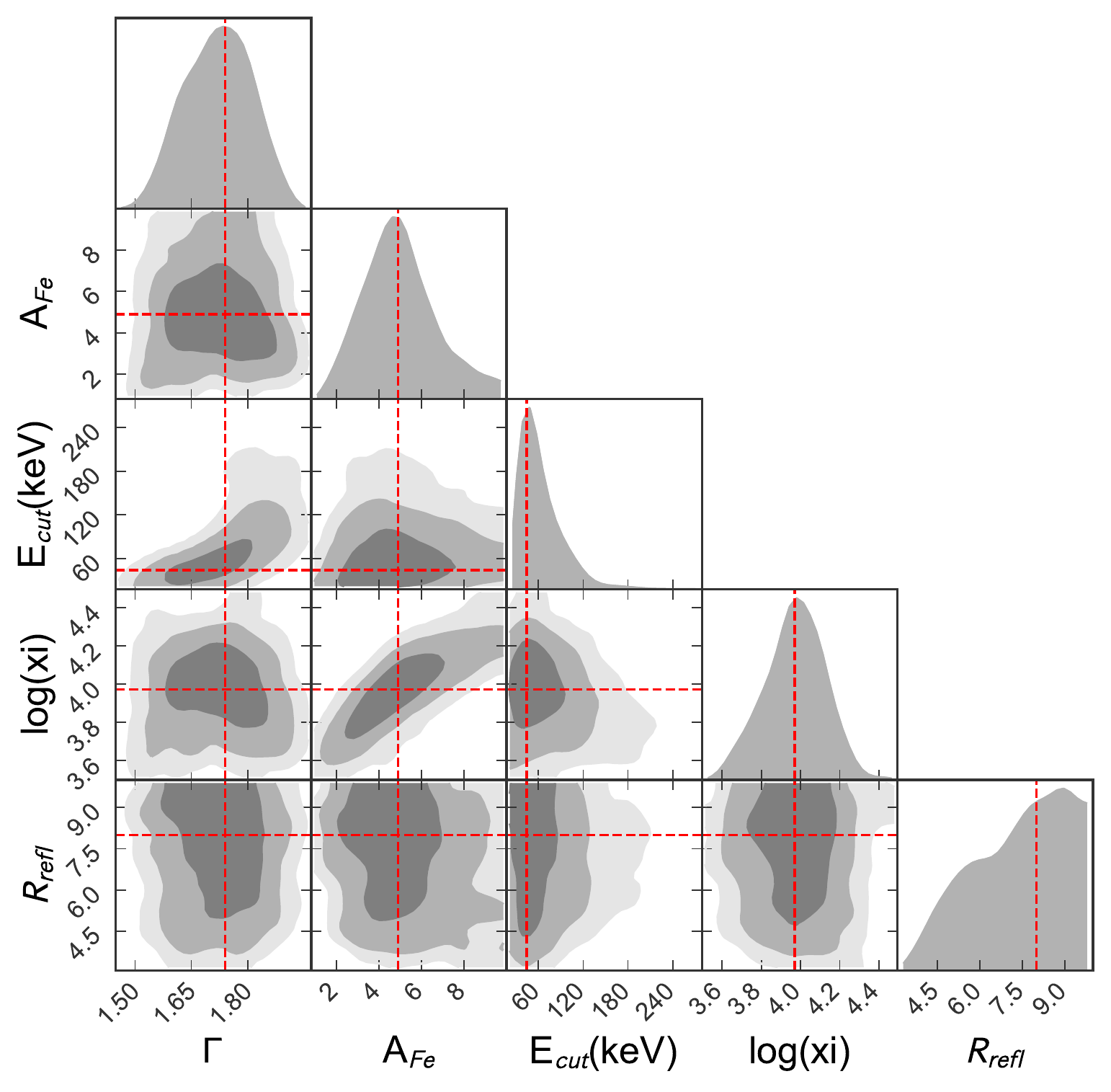}
	\includegraphics[scale=0.372]{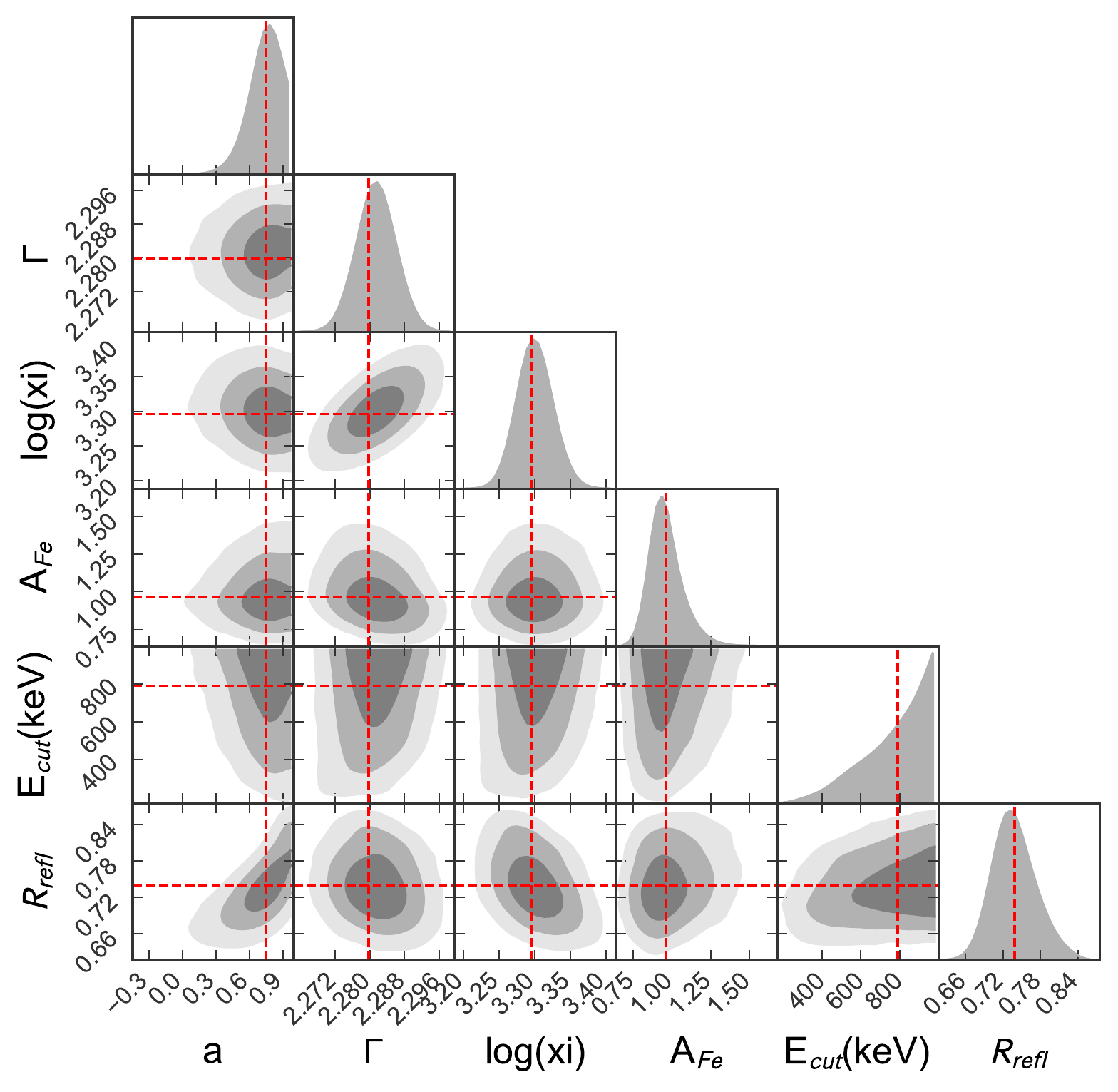}
	\includegraphics[scale=0.372]{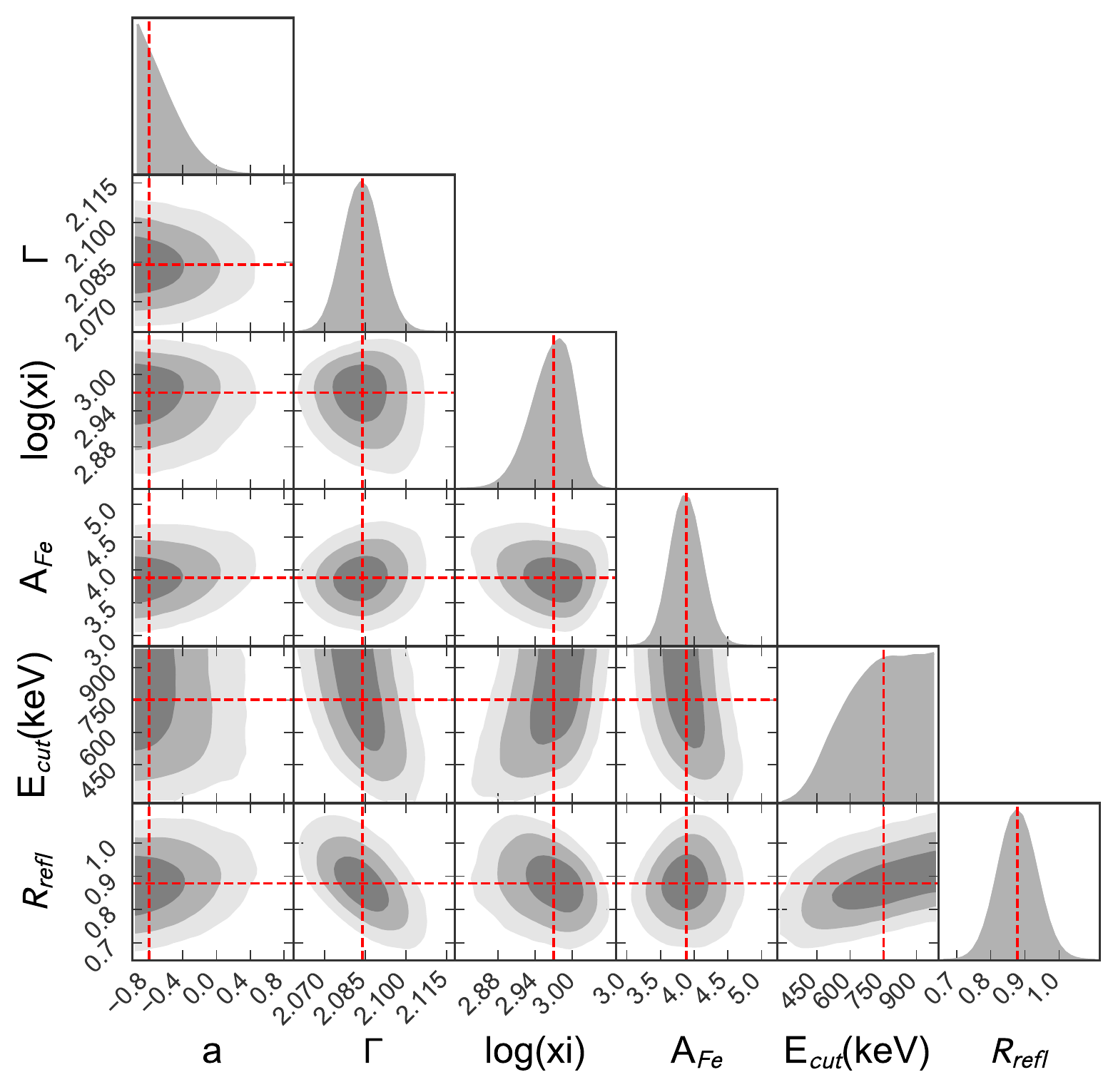}
	\includegraphics[scale=0.372]{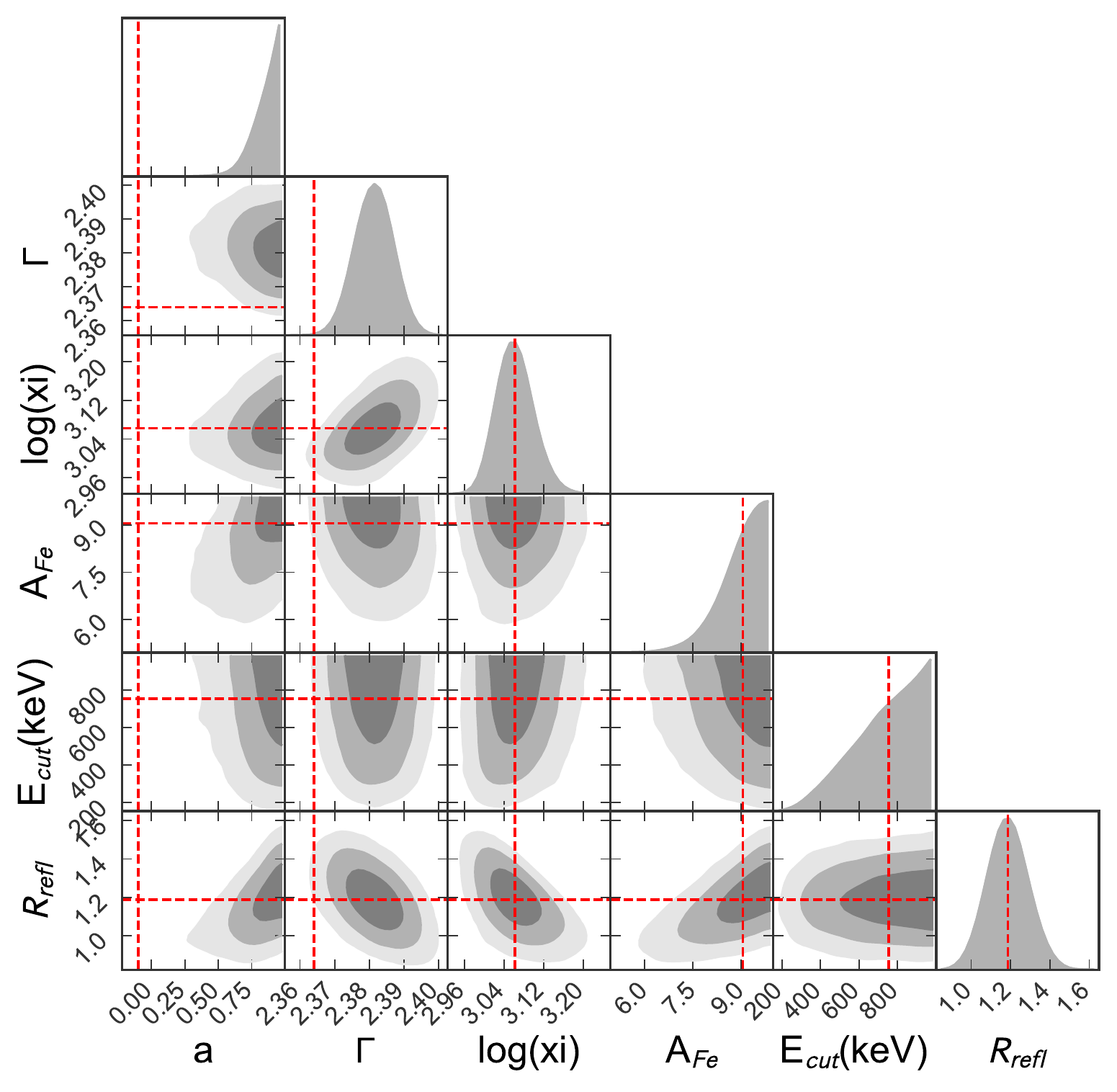}
	\includegraphics[scale=0.372]{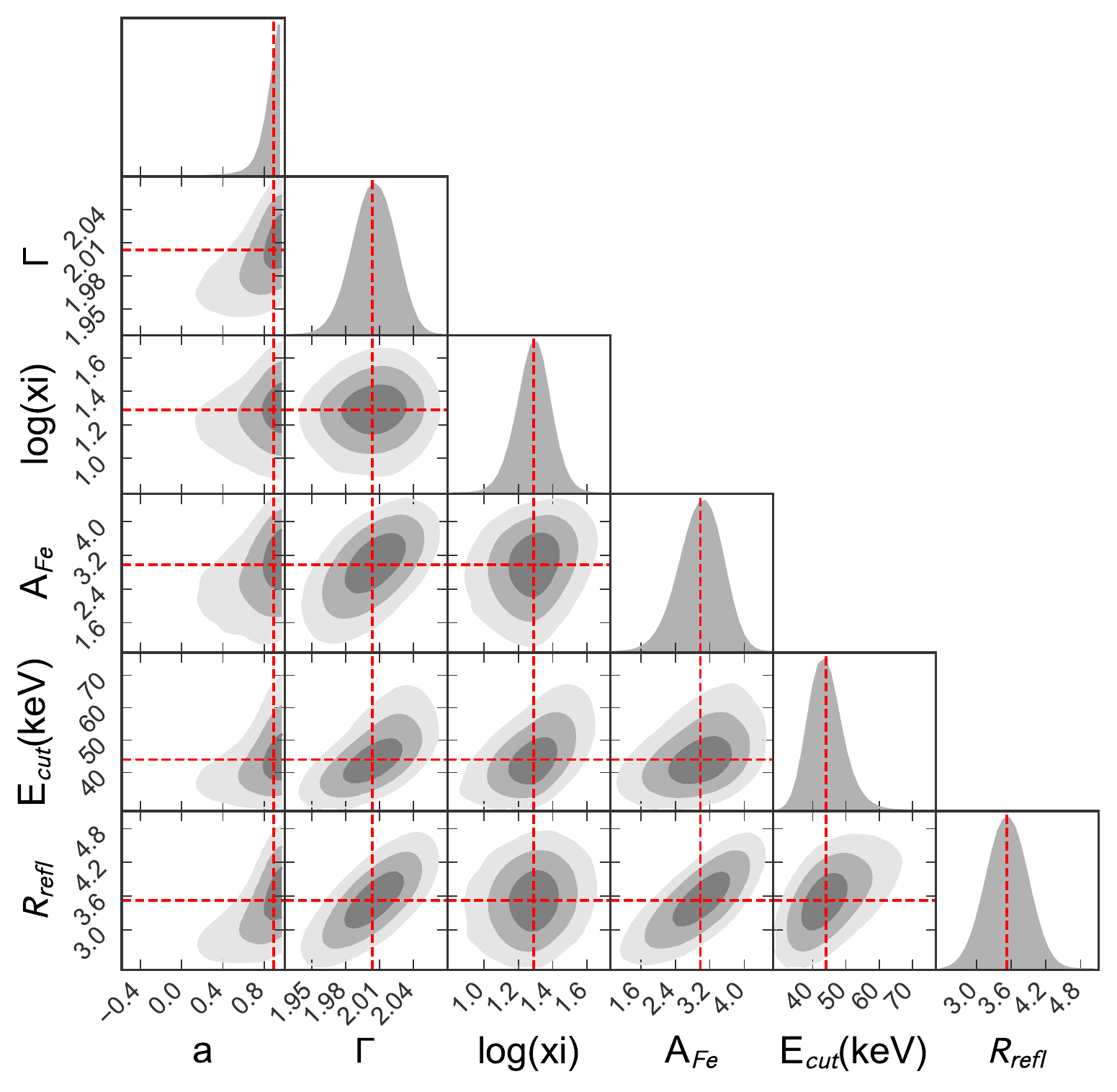}
	\includegraphics[scale=0.372]{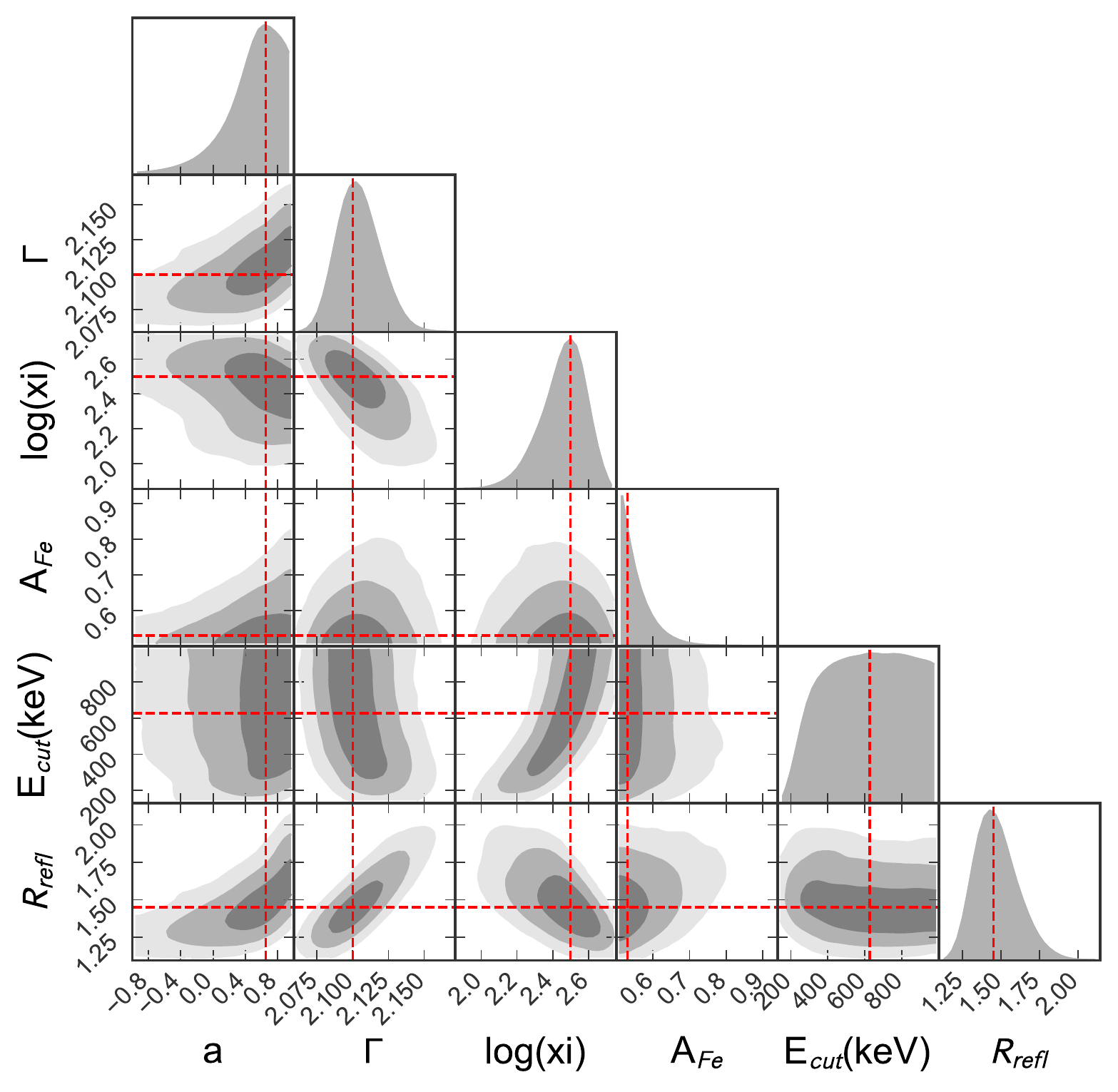}
	\includegraphics[scale=0.372]{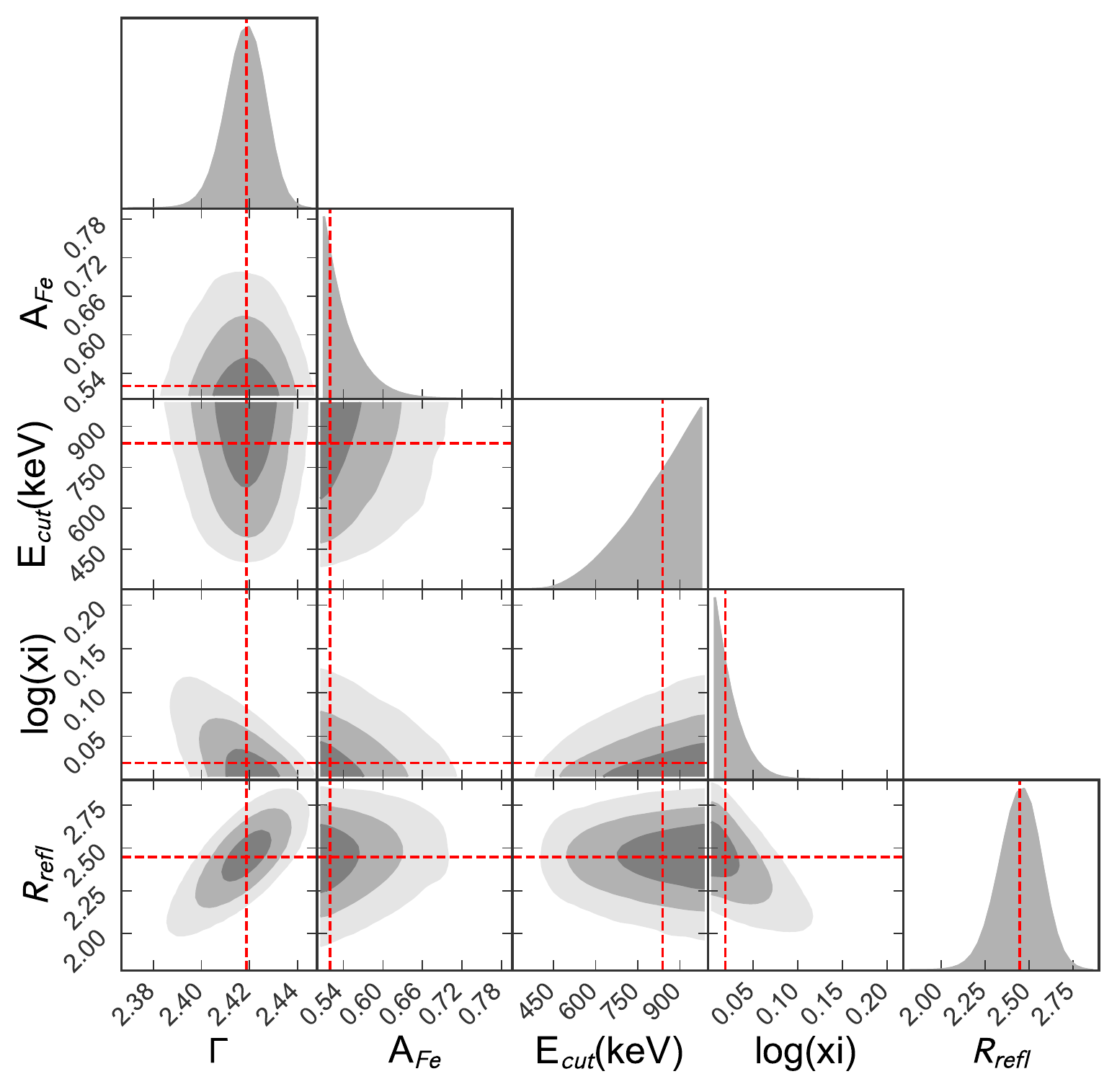}
	\includegraphics[scale=0.372]{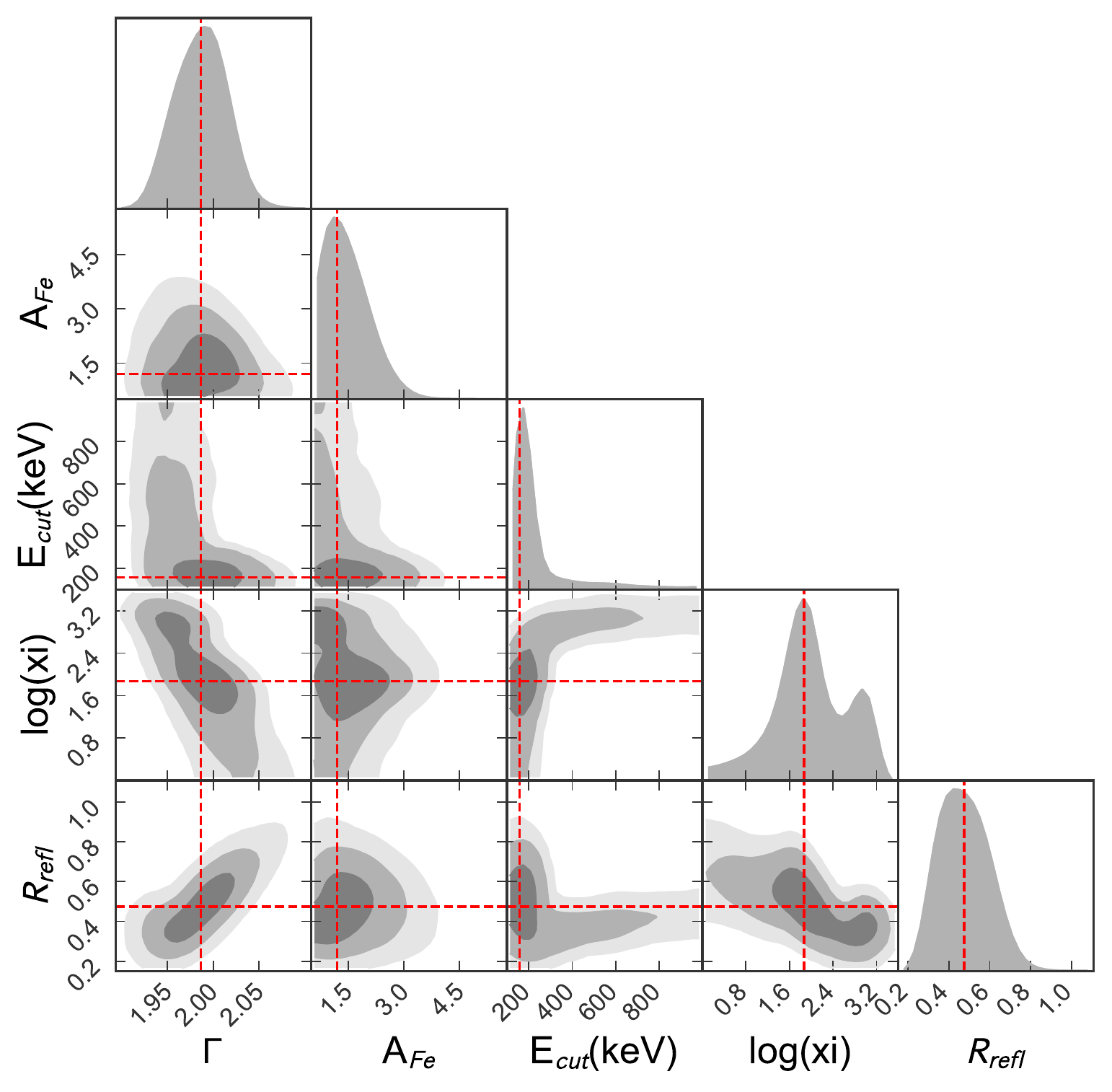}
    \caption{68\%, 90\% and 99\% contour plots resulting from the MCMC analysis of the best fitting models (see Table \ref{tab:best-fitting-param}) applied to the broad-band spectra of the super-Eddington sources IRAS\,04416+1216, IRASF\,12397+3333, Mrk\,493 (top panel from left to right) Mrk\,142, Mrk\,382,   PG\,0026+129 (central panel from left to right) PG\,0953+414 ans=d NGC\,4748 (bottom panel from left to right). We show the outputs for photon index ($\Gamma$), iron abundance ($A_{\rm Fe}$), cut-off energy ($E_{\mathrm{cut}}$[keV]), ionization parameter ($\log(\xi /\rm erg\,\rm cm\,\rm s^{-1}$)), reflection fraction ($R_{\rm refl}$) and black hole spin ($\mathit{a}$).}
    \label{fig:triangular}
\end{figure*}
%%%%%%%%%%%%%%%%%%%%%%%%%%%%%%%%%%%%%%%%%%%%%%%%%%

% Don't change these lines
\bsp	% typesetting comment
\label{lastpage}
\end{document}